\documentstyle[ukps1]{mn}

\newif\ifAMStwofonts
\def\der{{\rm d}} 			% roman derivative
\def\dd{{\rm d}} 			% roman derivative
\newcommand{\msun}{M_{\odot}}
\newcommand{\rsun}{R_{\odot}}
\newcommand{\rr}{R_{\rm R}}
\newcommand{\za}{\zeta_{\rm ad}}
\newcommand{\zr}{\zeta_{\rm R}}
\newcommand{\mej}{\Delta M_{\rm ej}}
\newcommand{\mig}{\Delta M_{\rm ig}}

\ifoldfss
  \ifCUPmtlplainloaded \else
    \NewTextAlphabet{textbfit} {cmbxti10} {}
    \NewTextAlphabet{textbfss} {cmssbx10} {}
    \NewMathAlphabet{mathbfit} {cmbxti10} {} % for math mode
    \NewMathAlphabet{mathbfss} {cmssbx10} {} %  "   "    "
  \fi
  \ifAMStwofonts
    \ifCUPmtlplainloaded \else
      \NewSymbolFont{upmath} {eurm10}
      \NewSymbolFont{AMSa} {msam10}
      \NewMathSymbol{\upi}     {0}{upmath}{19}
      \NewMathSymbol{\umu}     {0}{upmath}{16}
      \NewMathSymbol{\upartial}{0}{upmath}{40}
      \NewMathSymbol{\leqslant}{3}{AMSa}{36}
      \NewMathSymbol{\geqslant}{3}{AMSa}{3E}

    \fi
  \fi
\fi % End of OFSS

\ifnfssone
  \newmathalphabet{\mathit}
  \addtoversion{normal}{\mathit}{cmr}{m}{it}
  \addtoversion{bold}{\mathit}{cmr}{bx}{it}
  \newmathalphabet{\mathbfit} % math mode version of \textbfit{..}
  \addtoversion{normal}{\mathbfit}{cmr}{bx}{it}
  \addtoversion{bold}{\mathbfit}{cmr}{bx}{it}
  \newmathalphabet{\mathbfss} % math mode version of \textbfss{..}
  \addtoversion{normal}{\mathbfss}{cmss}{bx}{n}
  \addtoversion{bold}{\mathbfss}{cmss}{bx}{n}
  \ifAMStwofonts
    \ifCUPmtlplainloaded \else
      \UseAMStwoboldmath
      \makeatletter
      \new@mathgroup\upmath@group
      \define@mathgroup\mv@normal\upmath@group{eur}{m}{n}
      \define@mathgroup\mv@bold\upmath@group{eur}{b}{n}
      \edef\UPM{\hexnumber\upmath@group}
      \new@mathgroup\amsa@group
      \define@mathgroup\mv@normal\amsa@group{msa}{m}{n}
      \define@mathgroup\mv@bold\amsa@group{msa}{m}{n}
      \edef\AMSa{\hexnumber\amsa@group}
      \makeatother
      \mathchardef\upi="0\UPM19
      \mathchardef\umu="0\UPM16
      \mathchardef\upartial="0\UPM40
      \mathchardef\leqslant="3\AMSa36
      \mathchardef\geqslant="3\AMSa3E
    \fi
  \fi
\fi % End of NFSS release 1

\ifnfsstwo
  \DeclareMathAlphabet{\mathbfit}{OT1}{cmr}{bx}{it}
  \SetMathAlphabet\mathbfit{bold}{OT1}{cmr}{bx}{it}
  \DeclareMathAlphabet{\mathbfss}{OT1}{cmss}{bx}{n}
  \SetMathAlphabet\mathbfss{bold}{OT1}{cmss}{bx}{n}
  \ifAMStwofonts
    \ifCUPmtlplainloaded \else
      \DeclareSymbolFont{UPM}{U}{eur}{m}{n}
      \SetSymbolFont{UPM}{bold}{U}{eur}{b}{n}
      \DeclareSymbolFont{AMSa}{U}{msa}{m}{n}
      \DeclareMathSymbol{\upi}{0}{UPM}{"19}
      \DeclareMathSymbol{\umu}{0}{UPM}{"16}
      \DeclareMathSymbol{\upartial}{0}{UPM}{"40}
      \DeclareMathSymbol{\leqslant}{3}{AMSa}{"36}
      \DeclareMathSymbol{\geqslant}{3}{AMSa}{"3E}
    \fi
  \fi
\fi % End of NFSS release 2

\ifCUPmtlplainloaded \else
  \ifAMStwofonts \else % If no AMS fonts
    \def\upi{\pi}
    \def\umu{\mu}
    \def\upartial{\partial}
  \fi
\fi

\title{Properties of discontinuous and nova--amplified mass transfer
in CVs}  

\author[K.~Schenker et al.]{K.~Schenker$^{1,3}$, U.~Kolb$^{2,3}$, 
                            H.~Ritter$^{3}$ \\
           $^1$ Astronomisches Institut der Universit\"at Basel, 
                Venusstr.~7, CH-4102 Binningen, Switzerland \\
           $^2$ Astronomy Group, University of Leicester, Leicester
                LE1 7RH, U.K.\\
           $^3$ Max-Planck-Institut f\"{u}r Astrophysik,
	       Karl-Schwarzschild-Str.~1,
               D-85740 Garching, Germany }
\date{Accepted. Received}
\pagerange{\pageref{firstpage}--\pageref{lastpage}}
\pubyear{1998}

\begin{document}

\label{firstpage}

\maketitle

%------------------------------------------------------------
%============================================================
%------------------------------------------------------------

\begin{abstract}
We investigate the effects of discontinuous mass loss in
recurrent outburst events on the long--term evolution of cataclysmic
variables (CVs). Similarly we consider the effects of frictional angular
momentum loss (FAML), i.e.~interaction of the expanding nova envelope
with the secondary.
The Bondi--Hoyle accretion model is used to parameterize FAML in
terms of the expansion velocity $v_{\rm exp}$ of the nova envelope at
the location of the secondary; we find that
small $v_{\rm exp}$ causes strong FAML.

Numerical calculations of CV evolution over a wide range of parameters
demonstrate the equivalence of a discontinuous sequence of nova
cycles and the corresponding mean evolution (replacing 
envelope ejection by a continuous wind), even close to mass
transfer instability.
A formal stability analysis of discontinuous mass transfer 
confirms this, independent of details of the FAML model. 

FAML is a consequential angular momentum loss which amplifies the mass
transfer rate driven by systemic angular momentum losses such as
magnetic braking. We show that for a given $v_{\rm exp}$ and white
dwarf mass the amplification increases with secondary mass and 
is significant only close to the largest secondary mass consistent
with mass transfer stability. 
The amplification factor is independent of the envelope mass ejected
during the outburst, whereas the mass transfer amplitude induced by
individual nova outbursts is proportional to it.

In sequences calculated with nova model parameters taken
from Prialnik \& Kovetz 
\cite{prialnik:kovetz} FAML amplification is negligible,
but the outburst amplitude in systems below the
period gap with a white dwarf mass $\simeq 0.6 \msun$ is larger than a
factor of 10. The mass transfer rate in such systems is smaller than
$10^{-11} \msun$/yr for $\simeq 0.5$~Myr ($\simeq 10\%$ of the nova
cycle) after the outburst. This offers an explanation for
intrinsically unusually faint CVs below the period gap.
\end{abstract}

\begin{keywords}
novae, cataclysmic variables --- binaries: close --- stars:
evolution 
\end{keywords}

% ------------- Begin of Main Text -------------------------

\section{Introduction}
%***********************************************************

Cataclysmic variables (CVs) are short--period binary systems
in which a
Roche--lobe filling low--mass main--sequence secondary transfers
mass to a white dwarf (WD) primary. The
transferred matter accretes onto the WD either through a disc or a
stream and slowly builds up a hydrogen--rich surface layer on the WD. 
With continuing accretion the pressure at the bottom of this layer
increases, and hydrogen burning eventually starts. 
The thermodynamic conditions at ignition determine how the burning
proceeds (e.g.\ Fujimoto 1982). If the degeneracy is very high, a
thermonuclear runaway occurs, leading to a violent outburst
terminated by the ejection of all or most of the accumulated
envelope. Classical novae are thought to be objects undergoing such
an outburst (cf.\ Livio 1994 for a recent review).
Ignition at moderate or weak degeneracy causes strong or weak H
shell flashes, whereas stable or stationary hydrogen burning requires
fairly high accretion rates $\ga 10^{-7} \msun/$yr which are not
expected to occur in CVs.

Mass transfer in CVs is driven by orbital angular momentum losses 
which generally shrink the binary and maintain the semi--detached
state. The observed properties of short--period CVs below the CV 
period gap (orbital period $P \la 2$~h) are consistent with 
gravitational wave radiation as the only driving mechanism.
A much stronger angular momentum loss, usually assumed to be magnetic
stellar wind braking, is needed for systems above the
gap ($P \ga 3$~h). The assumption that 
magnetic braking ceases to be effective once the secondary becomes
fully convective in turn provides a natural explanation for the period
gap as a period regime where the systems are 
detached  and therefore unobservable (Spruit \& Ritter 
1983, Rappaport et al.\ 1983). 
The resulting typical mass transfer
rate $X$ in CVs is $X \simeq 5 \times 10^{-11} \msun/$yr below the gap and
$X \simeq 10^{-9} - 10^{-8} \msun/$yr above the gap (see e.g.\ King 
1988, Kolb 1996, for reviews).

For negligible wind losses from the system the accretion rate is
essentially the same as the transfer rate. With the above typical values 
the H ignition on the WD turns out to be degenerate enough
to cause more or less violent outbursts (e.g.\ Prialnik \& Kovetz 1995). 
As the mass to be accumulated before ignition is very
small ($\mig \simeq 10^{-6} - 10^{-3} \msun$) the outbursts recur on
a time $t_{\rm rec} = \mig/X$ much shorter than the mass transfer
timescale which determines the 
long--term evolution. Studies of the secular evolution of CVs make use
of this fact and replace a sequence of nova outbursts with given
recurrence time $t_{\rm rec}$ and ejected envelope mass $\mej$ 
by a continuous isotropic wind loss from the WD at a constant rate 
$\mej/t_{\rm rec}$.

Such a procedure obviously neglects any effect that nova outbursts may
have on the long--term evolution. These are in particular \\
(i)~The evolution of the system is not continuous but
characterized by sudden changes of the orbital parameters, causing
the mass transfer rate to fluctuate around the continuous wind average
value (see Sect.~2). It is not a priori clear if the continuous
wind average properly describes the system's evolution close to mass
transfer instability. \\
(ii)~At visual maximum (and the following decline) nova envelopes 
have pseudo--photospheric radii of typical giants, i.e.~much larger than
the orbital separation. Therefore the secondary is engulfed in this
envelope and possibly interacting with it. Drag forces on the 
secondary moving within the envelope can lead to frictional angular
momentum loss (FAML) from the orbit, and accretion of envelope
material onto the secondary could increase its photospheric metal
abundances (pollution). \\  
(iii)~The H burning hot WD is extremely luminous
($\sim 10^{\rm 4} L_{\odot}$, compared with
accretion luminosities $\sim 1 L_{\odot}$) for as long as
a few years. This might drive additional mass loss from the secondary
star. 

In this paper we will focus on the effects of mass loss
discontinuities and FAML. We neglect irradiation as it does not last long
enough to influence the long--term evolution.
Pollution is expected to be important only for metal poor secondaries
(Stehle 1993); we neglect it altogether.

In Sect.~2 we formally derive the continuous wind average and
consider the stability of mass transfer in the presence of nova
discontinuities analytically, with FAML of arbitrary strength. 
We review previous studies on FAML and follow Livio et al.\ \cite{livio:etal} 
to derive a simple quantitative model for FAML in Sect.~3.  
Using this description we perform numerical calculations of the
long--term evolution of CVs with various strengths of FAML, both for
sequences of nova outbursts and the continuous wind average. 
Results of such computations where the
FAML strength and the ejected mass per outburst have been varied
systematically are shown in Sect.~4. 
Sequences with FAML parameters taken from 
the consistent set of nova models by Prialnik \& Kovetz
\cite{prialnik:kovetz} are 
shown 
at the end of Sect.~4. Section 5 discusses our results.

\section{Mass transfer stability and classical novae} 
%***********************************************************

We begin by investigating how mass transfer discontinuities
induced by nova 
outbursts affect mass transfer stability. Introducting the
well--known conservative mass transfer stability criterion we 
develop a formalism to extend its applicability to the discontinuous
case. The strength of FAML enters as a free parameter.

\subsection{Conservative and continuous CV evolution}
\label{sec-2.1}
%+++++++++++++++++++++++++++++++++

Following Ritter \cite{ritter} the mass transfer rate $X$ in a CV can be
approximated by  
\begin{equation}
 X \equiv  -\dot{M}_{\rm 2}=\dot{M}_{\rm 0}\;\exp\left(\frac{R_{\rm 2}-\rr}
	{H_{\rm p}}\right) \, ,
 \label{eq-2.1.2}
\end{equation} 
(note that $X$ is always positive). Here both 
$\dot{M}_{\rm 0}$ $\simeq$ \mbox{$10^{\rm -8} M_{\odot}\:{\rm yr}^{\rm -1}$} 
and the ratio $\epsilon = H_{\rm p}/R_2 \simeq H_{\rm p}/R_{\rm R} \simeq 10^{\rm -4}$ 
of the photospheric pressure scale height
$H_{\rm p}$ to the donor's radius $R_2$ are roughly 
constant for the secondaries under
consideration. The secondary's Roche radius $\rr$ can be written as a
fraction $f_2$ of the orbital distance $a$, 
\begin{equation}
 \rr = f_{\rm 2}(q) \: a \, .
\label{eq-2.1.3}
\end{equation}
$f_2$ depends only on the mass ratio $q=M_1/M_2$ and is 
for $q>1.25$ given by  
$f_{2}(q) \simeq (8/81(1+q))^{1/3}$ to better than $2\%$
(Paczy\'{n}ski 1971).

\begin{figure}
\centerline{\plotone{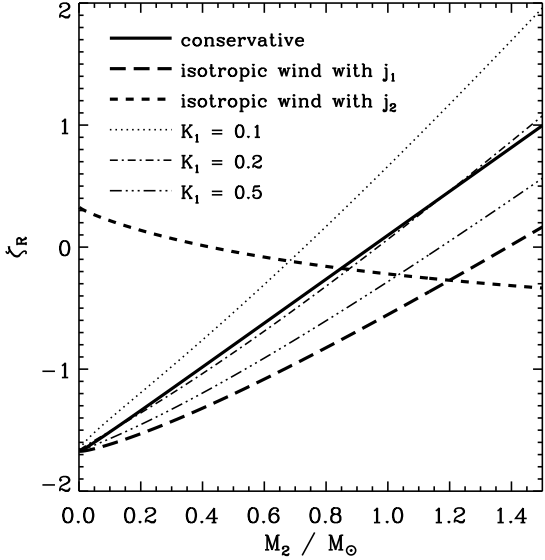}}
\caption[ ] 
{Reaction of the secondary's Roche radius to mass transfer and mass
loss, as a function of $M_2$ for fixed $M_{\rm 1}=1.2\:M_{\rm
\odot}$. Plotted is the mass--radius exponent for  
conservative mass transfer ($\zr^c$; full line), isotropic wind with 
$\eta = 0$, $\nu = 1/q$ ($\zr$ from (\protect\ref{eq-2.1.14}); long
dashed), and isotropic wind with $\eta = 0$, $\nu = q$ (short dashed). 
Also shown is $\zr^F$ for various values of $K_1$ (see Sect.~3.1).}
\label{fig:zetaroche}
\end{figure}

To assess how $X$ changes with time and 
to find a stationary value for $X$ (where $\dot X = 0$) 
we have to consider both $\dot R_2$ and $\dot \rr$. From the 
total orbital angular momentum $J$,
\begin{equation}
 J = M_{\rm 1} M_{\rm 2} \sqrt{ \frac{G a}{M} }
  \: ,
 \label{eq-2.1.1}
\end{equation}
and the derivative $\beta_2 = \dd \ln f_2 / \dd \ln q$ ($\simeq -
q/3(1+q)$) we find with (\ref{eq-2.1.3})
\begin{equation}
  \frac{\dot \rr}{\rr} = 
  (\beta_2 - 2) \frac{\dot M_1}{M_1} - (\beta_2
  + 2) \frac{\dot M_2}{M_2} + \frac{\dot M}{M} + 2 \frac{\dot J_{\rm
  m} + \dot J_{\rm sys}}{J} \, ,
\label{eq-2.1.5}
\end{equation} 
where $M_1$, $M_2$ and $M=M_1+M_2$ denote the primary mass, donor
mass and total mass, respectively.
In (\ref{eq-2.1.5}) we formally separate ``mass--loss related'' angular
momentum loss $\dot J_{\rm m}$, caused by mass which leaves the binary
and carries a certain specific angular momentum $j = \nu (J/M)$ 
(quantified by the dimensionless parameter $\nu$), i.e.\
\begin{equation}
  \dot J_{\rm m} = j \dot M = \nu \frac{J}{M} \dot M \, ,
\label{eq-2.1.6}
\end{equation}
and ``systemic'' angular momentum loss $\dot J_{\rm sys}$ which
operates without (noticeable) mass loss, e.g.\ gravitational wave
radiation and magnetic braking. By combining the mass--changing terms 
(\ref{eq-2.1.5}) is usually rewritten as 
\begin{equation}
  \frac{\dot \rr}{\rr} = \zr \frac{\dot M_2}{M_2} + 2
  \frac{\dot J_{\rm sys}}{J} \: ,
\label{eq-2.1.7}
\end{equation}
thus defining the mass--radius exponent $\zr$ of the
secondary's Roche radius. 
In the simple case of conservative mass transfer where
$M=$~constant, i.e.\ $\dot M_2 = - \dot M_1$, we find from 
(\ref{eq-2.1.5}) 
\begin{equation}
  \zr^c = 2 \left( \frac{M_2}{M_1} - 1 \right) - \frac{M}{M_1} \beta_2
        \: \simeq \: \frac{2}{q}  - \frac{5}{3} \: .
\label{eq-2.1.13}
\end{equation}
In the more general case of an isotropic wind loss at a 
rate $\dot M = (1-\eta) \dot M_2$, i.e.\ $\dot M_1 = -\eta \dot M_2$, 
we obtain 
\begin{equation}
  \zr = (1-\eta)\frac{2\nu+1}{1+q} + \eta\frac{2}{q} - 2 -
  \beta_2 \left( 1 + \frac{\eta}{q} \right) \: ,
\label{eq-2.1.14}
\end{equation}
an expression especially useful when $\eta$ and $\nu$ are
constant. Figure~\ref{fig:zetaroche} depicts $\zr^c$ and $\zr$ with
$\eta = 0$ and either $\nu=1/q$ or $\nu=q$ as a function of $M_2$ for
$M_1=1.2\msun$.  

\begin{figure}
\centerline{\plotone{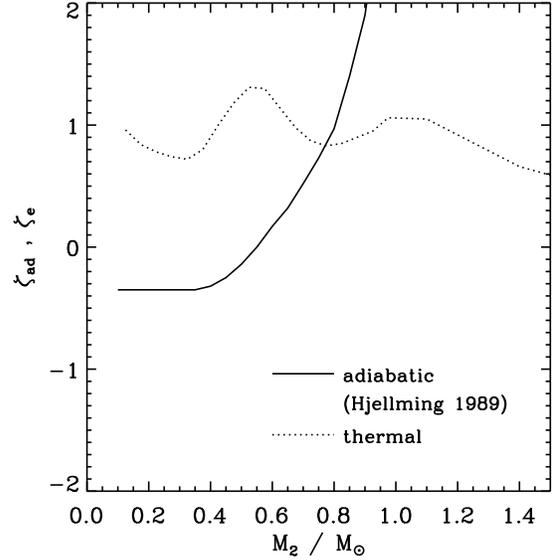}}
\caption[ ] 
{Reaction of the secondary's radius to mass loss
as a function of stellar mass $M_2$. Full line: adiabatic mass--radius
index $\za$; dotted line: thermal mass--radius
index $\zeta_{\rm e}$.
}
\label{fig:zetastar}
\end{figure}

Similarly, it is standard practice to decompose the secondary's
radius change into the adiabatic response $\za \dot M_2/M_2$ and the
thermal relaxation $K \equiv (\partial \ln R_2 /\partial t)_{M={\rm
const.}}$, 
\begin{equation}
  \frac{\dot R_2}{R_2} = \za \frac{\dot M_2}{M_2} + K \: .
\label{eq-2.1.8}
\end{equation}
Fig.~\ref{fig:zetastar} shows the adiabatic mass--radius exponent
$\za$ as a function of stellar mass for low--mass ZAMS secondaries
(Hjellming 1989). 
Fully convective stars and stars with deep convective envelopes ($M_2
\la 0.5 \msun$) have $\za \simeq -1/3$. 

Using these definitions the time derivative of $X$ according to
(\ref{eq-2.1.2}) becomes
\begin{equation}
  \dot X = X \: \left[{\cal A} - {\cal B} X \right] \: ,
 \label{eq-2.1.9}
\end{equation}
with 
\begin{equation}
 {\cal A} = \frac{1}{\epsilon} \: 
             \left( K - 2 \frac{\dot J_{\rm sys}}{J} \right) 
 \label{eq-2.1.10}
\end{equation}
and 
\begin{equation}
 {\cal B} = \frac{1}{\epsilon M_2} \: \left( \za - \zr \right) \: . 
 \label{eq-2.1.11}
\end{equation}

The binary attempts to settle at the stationary mass
transfer rate  
\begin{equation}
 X_s = \frac{\cal A}{\cal B} = M_2 \: \frac{K - 2\dot J_{\rm
 sys}/J}{\za - \zr} 
\label{eq-2.1.15}
\end{equation}
($\dot X = 0$ for $X=X_s$). The stationary rate is {\em stable} if
$\partial \dot X / \partial X < 0$ at $X=X_s$, i.e.\ if the system
opposes any instantaneous perturbation in $X$. This translates into
${\cal B} > 0$, hence the familiar criterion
\begin{equation}
  \za - \zr > 0 \Leftrightarrow \mbox{mass transfer (dynamically)
  stable} \: .
\label{eq-2.1.12}
\end{equation}

A similar stability criterion, $\zeta_{\rm e} - \zr > 0$,
against thermal timescale mass transfer can be derived (see e.g.\
Ritter 1996), where 
$\zeta_{\rm e}$ is the thermal equilibrium mass--radius exponent
($\simeq 0.85$ for low--mass main--sequence stars, see 
Fig.~\ref{fig:zetastar}).

\subsection{Non--stationary mass transfer}
%+++++++++++++++++++++++++++++++++++++++++++++++++++++++++++

The differential equation (\ref{eq-2.1.9}) governs the evolution of the
mass transfer rate as a function of time $t$. For constant $\cal A$ and
$\cal B$ 
the solution of (\ref{eq-2.1.9}) is
\begin{equation}
  \frac{1}{X(t)} = \frac{1}{X_s} - \left( \frac{1}{X_s} -
  \frac{1}{X_0} \right) \exp \left( -{\cal A}t \right) \: ,
\label{eq-2.2.1}
\end{equation}
where $X_0 \equiv X(t=0)$ denotes the initial value of $X$.
Hence any changes of the transfer rate proceed on the
characteristic timescale $t_c = 1/{\cal A} \sim {\cal O}(\epsilon
t_J)$, i.e.\ $t_c$ is a small fraction $\epsilon$ of the systemic angular  
momentum loss timescale $t_J = -J/\dot J_{\rm sys}$. 

In reality both $\cal A$ and $\cal B$ change with time,  
typically on a the secular timescale $t_J$, but for most practical
cases where we investigate mass transfer stability against short timescale 
perturbations $\cal A$ and $\cal B$ can be considered as
constant.   

%In reality neither $\cal A$ nor $\cal B$ are constant but change  
%typically on a the secular timescale $t_J$. Hence for most practical
%cases where we investigate mass transfer stability against short timescale 
%perturbations $\cal A$ and $\cal B$ can indeed be considered as
%constant.   

%$K$ is of order $(L_{\rm g}/L_2)/t_{\rm ce}$,
%where $t_{\rm ce} = GM_2 M_{\rm ce}/R_2L_2$ is the thermal time of the
%donor's convective envelope (with mass $M_{\rm ce}$; $L_2$ is
%the luminosity of the secondary). Although $t_{\rm ce}$ is 
%short for high--mass donors with a small convective envelope ($t_{\rm
%ce} \simeq 10^5$~yr for $M_2 = 1.2 \msun$) it increases considerably 
%with decreasing secondary mass ($\log t_{\rm ce}/{\rm yr} \simeq 6, 7,
%8, 9$ for  $M_2/\msun = 1.0, 0.70, 0.45, 0.15\msun$, respectively).

Formally (\ref{eq-2.2.1}) is a solution of (\ref{eq-2.1.9}) 
even if ${\cal A}< 0$ (i.e.\ $\left|K\right| > \left|2 \dot J_{\rm sys} /
J\right|$). In this case, assuming dynamical stability
(${\cal B} > 0$), $X_s$ is negative and no longer a stationary value for the
mass transfer rate. Rather (\ref{eq-2.2.1}) shows that in this case
$X$ decreases exponentially, i.e.\ the system detaches (note that
$X_0$ is a physical mass transfer rate and as such always positive). 
If finally the mass transfer
is unstable (${\cal B} < 0$) then we see from (\ref{eq-2.1.9})
that unless $\cal A$ has a large negative value the transfer rate
grows. The growth timescale is initially $t_c$, but becomes shorter
and shorter with further increasing $X$. 
%this last sentence does already cover what you say in the next:
%If both ${\cal A}$
%and ${\cal B}$ are negative ${\cal A}$ competes with ${\cal B} X$
%for the sign of $\dot X$ (remember that $X>0$ always).

\subsection{Nova--induced discontinuities}
%++++++++++++++++++++++++++++++++++++++++
 
So far we considered only continuous (though not necessarily
stationary) mass transfer. Events that change orbital parameters on a
timescale much shorter than the characteristic time $\epsilon t_J \simeq
10^4$~yr for reestablishing the local stationary value $X_s$ may be
regarded as discontinuous and instantaneous.
Nova outbursts certainly belong to this category; the nova envelope
expands beyond the orbit within less than a few $10^2$~days after
ignition and returns within $1 - 10$~years (actually these are upper
limits for the slowest novae). 
In the following we apply the above formalism separately to the
outburst phase and the inter--outburst phase, then combine them 
to describe the full nova
cycle.

\subsubsection{Outburst phase}
%-------------------------
As a result of the outburst the envelope mass $\mej$ is ejected.
We expect that the specific angular momentum carried
away by $\mej$ is higher than the specific orbital angular momentum 
of the WD due to dynamical friction of the secondary orbiting
within the envelope (FAML).
Hence we write for the discontinuous change of
the orbital angular momentum
\begin{equation}
   \Delta J = - (\nu_1 + \nu_{\rm FAML}) \frac{J}{M} \mej \, ,
\label{eq-2.3.1}
\end{equation}
where $\nu_1 = M_2/M_1$ accounts for the specific orbital angular
momentum  
$j=j_1$ of the WD (represented as a point mass). The free parameter 
$\nu_{\rm FAML}$ measures the strength of FAML and will be estimated
in terms of a simple model for the frictional processes in Sect.~3
below. 

Correspondingly, using (\ref{eq-2.1.5}) with
$\Delta M = \Delta M_1 = - \mej$ and $\Delta M_2 = 0$
(i.e.\ neglecting the small amount of mass accreted onto the
secondary as well as any mass transfer during this short phase), 
the change $\Delta \rr = \rr (\mbox{post}) - \rr (\mbox{pre})$ 
of the Roche radius is  
\begin{equation}
  {\left( \frac{\Delta \rr}{\rr} \right)}_{\rm out} = 
  \frac{\mej}{M_1} \left( \frac{q}{1+q} - \beta_2
  - \frac{2q}{1+q} \nu_{\rm FAML} \right) \, .
\label{eq-2.3.2}
\end{equation}

\subsubsection{Inter--outburst phase}
%-------------------------------
After the outburst mass transfer continues. 
Once the mass accreted onto the WD exceeds 
the critical ignition mass $\mig > 0$ the next outburst occurs. 
For simplicity we assume {\em conservative} mass transfer during 
the inter--outburst phase.
According to (\ref{eq-2.1.7}) the total change of the Roche radius in
the inter--outburst phase is
\begin{equation}
  {\left( \frac{\Delta \rr}{\rr} \right)}_{\rm inter} = 
   \zr^c \frac{- \mig}{M_2} + 2 \frac{\Delta J_{\rm sys}}{J} \: ,
\label{eq-2.3.4}
\end{equation}
where $\Delta J_{\rm sys}$ is the total
change of the orbital a.m.\ due to systemic losses between outbursts.
%The total change of the Roche radius
%over a complete cycle then is $\Delta
%\rr (\mbox{total}) = \Delta \rr (\mbox{out}) + \Delta \rr(\mbox{inter})$.

\subsubsection{Combined description}
%-------------------------------
For the study of the long--term evolution of CVs it is convenient to
replace the sequence of nova cycles with mass transfer discontinuities
by a mean evolution where the mass $\mej$ is regarded as being lost
continuously at a constant rate over the cycle in form of an isotropic stellar
wind carrying the specific orbital angular momentum $\left(\nu_1 +
\nu_{\rm FAML}\right) J/M$. If $\zr^F$ denotes the corresponding 
Roche lobe index then we have from (\ref{eq-2.1.7}) 
\begin{equation}
  {\left( \frac{\Delta \rr}{\rr} \right)}_{\rm total} = 
   \zr^F \frac{\Delta M_2}{M_2} + 2 \frac{\Delta J_{\rm
   sys}}{J} \: , 
\label{eq-2.3.5}
\end{equation}
again with $\Delta M_2 = - \mig$ because we neglect any change of $M_{2}$ 
during outburst. As the total change of the Roche radius
over a complete cycle is $\Delta
\rr (\mbox{total}) = \Delta \rr (\mbox{out}) + \Delta \rr(\mbox{inter})$,
comparison with (\ref{eq-2.3.2}) and (\ref{eq-2.3.4}) gives 
\begin{equation}
  \zr^F = \zr^c + \frac{ {\left( \Delta \rr/\rr \right)}_{\rm
  out}}{-\mig / M_2} \: .
\label{eq-2.3.6}
\end{equation}
This can be written as
\begin{eqnarray}
\label{eq-2.3.7}
  \zr^F & = & \zr^c - \frac{1-\eta^n}{q} \left( \frac{q}{1+q} - \beta_2  
    - \frac{2 \: q }{1+q} \nu_{\rm FAML} \right) 
 \\ \nonumber 
 & \simeq &  \zr^c - \frac{1-\eta^n}{1+q} \left( \frac{4}{3} - 2 \: 
 \nu_{\rm FAML} \right) \: .
\end{eqnarray}
Here $1 - \eta^n = \Delta M_1/\Delta M_2 = \mej/\mig$ specifies the
change of the WD mass {\em during outburst} in units of the mass lost from $M_2$
{\em during the inter--outburst phase}.
Equation (\ref{eq-2.3.7}) in fact is equivalent
to (\ref{eq-2.1.14}) with $\eta = \eta^n$ and $\nu=\nu_1+\nu_{\rm FAML}$;   
$\eta^n$ determines if the WD mass will grow or shrink in the
long--term evolution.

The generalization of (\ref{eq-2.3.7}) which allows for non--conservative
mass transfer between outbursts and a change of $M_2$ during outburst
is given in the Appendix.

\subsection{Stability of discontinuous mass transfer}
\label{sec-2.4}
%++++++++++++++++++++++++++++++++++++++++++++++++++++

As was shown in Sect.~2.1 mass transfer in the fictious
mean evolution which mimics the effect of nova outbursts by a
continuous wind loss is dynamically stable if   
\begin{equation}
 \za - \zr^F > 0 \: .
\label{eq-2.4.1}
\end{equation}
Clearly, in the case of discontinuous mass transfer with nova cycles
the stability considerations leading to this criterion
are no longer applicable, as the system never settles at a stationary
transfer rate. 
After an outburst the transfer rate $X$ follows 
the solution (\ref{eq-2.2.1}) and approaches the conservative
stationary rate $X_s^c = {\cal A}/{\cal B}^c$ (where ${\cal B}^c$ is
${\cal B}$ according to (\ref{eq-2.1.11}) with $\zr^c$).
At the next outburst $X$ changes discontinuously from the 
pre--outburst value $X_{\rm pre}$ to the (new) post--outburst value
$X_{\rm post}$, i.e.\ increases (or drops) by a factor  
\begin{equation} 
   \frac{X_{\rm post}}{X_{\rm pre}} = 
   \exp \left[ -\frac{1}{\epsilon} \left( \frac{\Delta \rr}{\rr}
   \right)_{\rm out} \right] \: ,
\label{eq-2.4.2}
\end{equation}
see Eqs.~(\ref{eq-2.1.2}) and (\ref{eq-2.3.2}).
In the following we consider mass transfer stability in the
presence of nova cycles and derive a generalized stability criterion
that reduces indeed to the simple form (\ref{eq-2.4.1}). 

In our approach we compare the mass transfer rate $X_{\rm pre}^i$ and
$X_{\rm pre}^{i+1}$ {\em immediately before}~subsequent outbursts $i$
and $i+1$ in  a sequence of cycles with constant $\mig$, $\mej$,
${\cal B}^c$ and $\cal A$. Eq.~(\ref{eq-2.2.1}) with $X_0 = X^i_{\rm
post}$ describes $X$ as a function of time between outbursts. 
At time $t=t_{\rm rec}$ when the outburst $i+1$ ignites,
the mass transferred since the last outburst is just $\mig$, i.e.\ 
\begin{equation}
 \int_{\rm 0}^{t_{\rm rec}} X(t) \: \der t = \mig \: .
\label{eq-2.4.10}
\end{equation}
This can be solved for the outburst recurrence time $t_{\rm rec}$
(cycle time)  
\begin{equation}
 t_{\rm rec} = \frac{1}{{\cal A}}\:\ln \left[ 1 + \frac
	{\exp\left({\cal B}^c\:\Delta M_{\rm ig}\right)-1}
	{X^i_{\rm post}/X_s^c} \right]
	\: .
 \label{eq-2.4.3}
\end{equation}
Inserting $t_{\rm rec}$ for $t$ in (\ref{eq-2.2.1}) gives the new
pre--outburst value $X^{i+1}_{\rm pre} = X(t_{\rm rec})$,
\begin{eqnarray}
\label{eq-2.4.4}
 X^{i+1}_{\rm pre} 
 =  X_s^c \left[ 1 - \exp (-{\cal B}^c\mig) \right] 
\\ \nonumber
  + X^i_{\rm pre} \exp \left( -{\cal B}^F \mig \right) \: ,
\end{eqnarray}
where we used (\ref{eq-2.4.2}) to replace $X^i_{\rm post}$ by
$X^i_{\rm pre}$,  
(\ref{eq-2.3.6}) and the definition of ${\cal B}^F$ analogous to
${\cal B}^c$, i.e.\ ${\cal B}^F = (\za - \zr^F)/\epsilon M_2$.
Hence the relative change of the pre--outburst value is 
\begin{eqnarray}
\label{eq-2.4.5}
 {\cal F} \lefteqn{ (X^i) = \frac{X^{i+1} - X^i}{X^i} = } 
 \\ \nonumber 
	  \lefteqn{
    \frac{X_s^c}{X^i} \left[ 1 - \exp (-{\cal B}^c\mig) \right] 
    - \left[ 1- \exp \left( -{\cal B}^F\mig \right) \right] }
 \\ \nonumber 
\lefteqn{
    \simeq \mig \left( \frac{{\cal A}}{X^i} - {\cal B}^F \right) 
 }
\end{eqnarray}
(the subscripts for all $X$ have been dropped as they all read
``pre''). We will refer to $\cal F$ as the growth function.

\begin{figure}
\centerline{\plotone{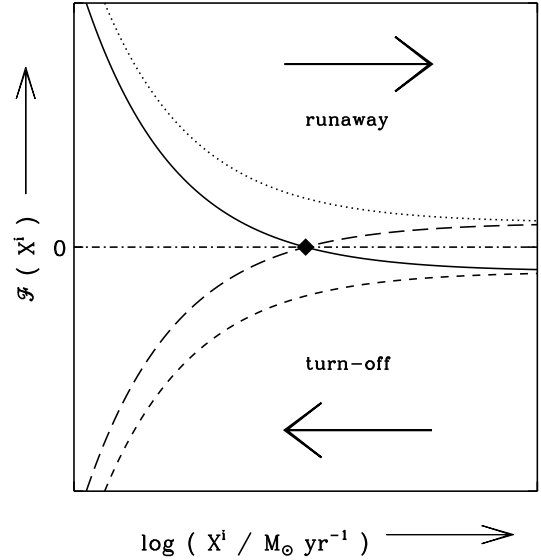}}
\caption[ ] 
{Growth  function ${\cal F}$,
defined in Eq.~(\protect\ref{eq-2.4.5}), as a function of 
pre--outburst mass transfer rate $X^i$ (schematically).
See Table~\ref{tab:fhat} for a description of the different curves.
Only the solid curve represents a stable stationary solution $\tilde{X}$ 
(marked by a filled diamond), all other cases lead to a runaway or a turn--off.}
\label{fig:fhat}
\end{figure}

For ${\cal F} > 0$ the mass transfer rate {\em grows} from outburst
to outburst, for ${\cal F} < 0$ it decreases, on a timescale
$t_{\rm rec}/{\cal F}$. Formally, ${\cal F} = 0$ for $X = \tilde{X}$, with 
\begin{equation}
   \tilde{X} = X_s^c \frac{1 - \exp \left( -{\cal B}^c \mig \right)}{1 -
   	\exp \left( -{\cal B}^F \mig \right)} \simeq 
	\frac{\cal A}{{\cal B}^F} \: .
\label{eq-2.4.6}
\end{equation}
$\tilde{X}$ represents the stationary {\em pre--outburst} mass 
transfer rate if it is positive, i.e.\ if $\cal A$ and ${\cal B}^F$
have the same sign. 
For it to be stable we require $\partial {\cal F}/\partial
X^i < 0$ at $X^i=\tilde{X}$ so that the system evolves back to $\tilde{X}$
after a perturbation increases/decreases $X$. This is the case if, and
only if, ${\cal A} > 0$. Then
the stationary solution exists only if ${\cal B}^F >0$, in all other
cases the system either detaches, or the transfer rate grows unlimited. 
Figure~\ref{fig:fhat} summarizes schematically the formal functional
dependency of $\cal F$ on $X^i$  for various combinations of the signs
of $\cal A$ and ${\cal B}^F$ (see Table~\ref{tab:fhat}). 
Hence the generalized stability criterion for discontinuous
nova--amplified mass transfer is ${\cal B}^F>0$, which is indeed
equivalent to (\ref{eq-2.4.1}). Remarkably,
this condition is independent of the sign of ${\cal B}^c$, i.e.\ the
system can evolve in a stable manner with formally dynamically
unstable conservative  mass transfer between subsequent outbursts.

\begin{table}
 \caption[ ]
 {Branches of possible solutions of Eq.~(\protect\ref{eq-2.4.5}) as
 shown in Fig.~\protect\ref{fig:fhat}.} 
 \label{tab:fhat}
 \begin{center}
 \begin{tabular}{c c c c}
 \hline
	 {\sc line} & \multicolumn{2}{c}{\sc sign of} &  {\sc stationary}  
 \\ \cline{2-3}
	 {\sc style} & ${\cal A}$ & ${\cal B}^F$ &  {\sc solution}
 \\ \hline
	full         & +  & +  & yes (stable) 
 \\
	long dashes  & -- & -- & yes (unstable) 
 \\
	dotted       & +  & -- & no
 \\
	short dashes & -- & +  & no 
 \\  \hline
 \end{tabular}
 \end{center}
\end{table}

The analysis of the system's behaviour once it violates condition
(\ref{eq-2.4.1}) is not straightforward. The main problem is that 
$\cal A$ is expected to change its sign (from positive to
negative) at about the same time as ${\cal B}^F$. 
This can be seen from an estimate
of the thermal relaxation term $K$ if the real evolution is replaced
by the continuous wind average: Stehle et
al.\ \cite{stehle:etal} have shown that the secular evolution rapidly converges to
an attracting evolutionary path characterized by 
\begin{equation}
  \frac{\dot R_2}{R_2} = \zeta_{\rm e} \frac{\dot M_2}{M_2}  \: ,
\label{eq-2.4.8}
\end{equation}
%where $\zeta_{\rm e}$ is the thermal equilibrium mass--radius exponent
%($\simeq 0.85$ for low--mass main--sequence stars), 
whatever the
initial configuration of the system. Assuming stationarity, 
equating (\ref{eq-2.4.8}) with (\ref{eq-2.1.7}), and
(\ref{eq-2.1.8}) with (\ref{eq-2.1.7}) gives after 
elimination of $\dot M_2/M_2$ 
\begin{equation}
  {\cal A} \simeq \frac{-2}{\epsilon} \frac{\dot J_{\rm sys}}{J} 
	\frac{\za - \zr^F}{\zeta_{\rm e} - \zr^F}  
\label{eq-2.4.9}
\end{equation}
as an estimate for $\cal A$. Hence ${\cal A} > 0$ during a phase of
stable, stationary mass transfer, but ${\cal A} < 0$ once $\za <
\zr^F$.
%%% (assuming $\zeta_{\rm e} > \za$).

In practice this means that when a stable system, characterized by the
full line in Fig.~\ref{fig:fhat}, approaches instability, the curve
becomes flatter and flatter with $\tilde{X}$ almost constant, and then
inverts to the (unstable) long--dashed line.

\section{A simple model for FAML}
%***********************************************************

In the previous section we derived analytic expressions describing
stationary mass transfer and mass transfer stability in the presence
of nova cycles. A non--zero FAML effect was allowed, and the
strength of FAML was treated as a free parameter $\nu_{\rm FAML}$. 
To proceed further and underline the above findings with numerical
examples we quantify $\nu_{\rm FAML}$ by adopting a model for
the frictional processes leading to FAML.

MacDonald \cite{macdonald} discussed 
for the first time the possibility that an interaction between the
secondary and the extended envelope might tap energy from the orbit
%(in analogy to the 'spiral--in' during the common envelope
%phase CV progenitors go through, cf.\ Paczy\'{n}ski \cite{paczynski2})
and thereby reduce the nova decay time. He 
determined the angular momentum transferred from the orbit to the
envelope by describing the nova 
envelope as a polytrope with index 3 at rest, and 
restricting the combined nuclear, internal and frictional energy
generation to the Eddington luminosity (MacDonald 1986).
Shara et al.\ \cite{shara:etal} and Livio et al.\ \cite{livio:etal} used a
more direct way, based on Bondi--Hoyle accretion, to estimate the
transfer of angular momentum. Their approach (which we adopt below) 
offers a very simple analytic treatment, as the explicit
expression for FAML has only one free parameter, 
the envelope expansion velocity $v_{\rm exp}$, 
measuring the strength of FAML. 
%Livio et al.~\cite{livio:etal} were 
%able to resolve a controversy on the direction of the change in
%orbital separation due to nova outbursts by pointing out the
%importance of this parameter and its dependence on the orbital period.
More recently, Kato \& Hachisu \cite{kato:hachisu} have again  
confirmed their earlier result that FAML has only minor influence 
on the decay phase of nova outbursts. 
This can be understood in terms of the high expansion velocities
in their models and will be discussed in Sect.~5. 
On the other hand, Lloyd et al.\ \cite{lloyd:etal} showed that common enevlope
evolution can contribute to the shaping of the nova remnant.

\subsection{FAML description according to Livio et al.~\protect\cite{livio:etal}} 
\label{sec-3.1}
%++++++++++++++++++++++++++++++++++++++++++++++++++++

In Bondi--Hoyle accretion linear momentum is transferred through
the drag force 
\begin{equation} 
 \vec{F}_{\rm drag} = - c_{\rm drag}({\cal M},\gamma_{\rm gas}) \: \frac{1}{2} \: \pi \:
	R_{\rm A}^{\rm 2} \: \rho \: v_{\rm rel} \vec{v}_{\rm rel} \: ,
 \label{eq:fdrag}
\end{equation}
where $\vec{v}_{\rm rel}$ is the gas stream velocity at infinity, $R_{\rm A}$ the
accretion radius (see below) and $\rho$ the density of the accreted
medium. The dimensionless drag coefficient $c_{\rm drag}$
varies with Mach number $\cal M$ and specific heat ratio $\gamma_{\rm gas}$
and is of order unity. Equation~(\ref{eq:fdrag}) 
is strictly valid for the highly supersonic case, 
but also applicable for ${\cal M} \ga 1$ if an adequate interpolation
for the accretion radius is used, e.g.\ 
\begin{equation} 
 R_{\rm A} = \frac{2\:G\:M_{\rm 2}}{v_{\rm rel}^{\rm 2}+c_{\rm S}^{\rm 2}}
 \label{eq:racc}
\end{equation}
(Shima et al.\ 1985);
$c_{\rm S}$ is the local sound speed. 
Detailed 3--dim.\ hydrodynamical calculations summarized in 
Ruffert \cite{ruffert} confirm that (\ref{eq:fdrag}) describes the 
subsonic case as well when $\pi R_{\rm A}^2$ is replaced by the
geometrical cross--section. But these simulations also reveal
additional complexities not represented in the simple form
(\ref{eq:fdrag}), e.g.\ the influence of the accretor's size, even in
the supersonic case. Most important, the drag coefficient can
deviate significantly from unity. Kley et al.~\cite{kley:etal}
point out that radiation pressure may reduce $c_{\rm drag}$ by a
factor of $\sim 20$ because it increases the sound speed and thus
lowers $\cal M$ to the subsonic case. 

Given these uncertainties, and the fact that idealized
Bondi--Hoyle accretion is merely a rough approximation to the
situation of a secondary orbiting in an expanding nova envelope, 
the quantitative expressions for FAML derived below are order of magnitude
estimates only. 

\begin{figure}
\centerline{\plotone{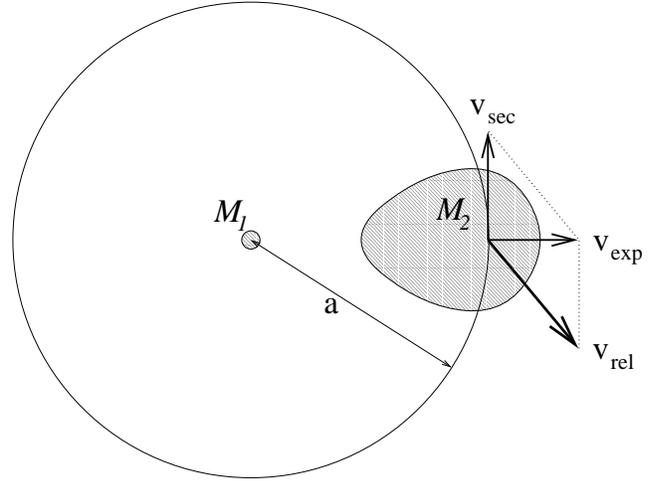}}
\caption[ ] 
{Sketch of the simplified geometry assumed to derive $\nu_{\rm FAML}$. 
The WD ($M_{\rm 1}$) is at rest and   
orbited by the secondary ($M_{\rm 2}$) at distance $a$ with 
velocity $v_{\rm sec}$. The envelope expands spherically symmetric
from the WD, and has constant expansion velocity $v_{\rm exp}$ at the
location of the secondary. }
\label{fig:sketch}
\end{figure}    

We now consider the FAML situation in a simplified geometry 
(Fig.~\ref{fig:sketch}): the WD is 
at rest in the centre of a spherically symmetrical expanding nova
envelope. Then the secondary's orbital velocity 
\begin{equation}
 v_{\rm sec} = 	\sqrt{\frac{G\left(M_{\rm 1}+M_{\rm 2}\right)}{a}}
 \label{eq:vsec}
\end{equation}
(typically $v_{\rm sec} \simeq 400-500$~km/s) is perpendicular to the 
wind expanding with $v_{\rm exp}$. Hence the velocity of the
accretion flow with respect to the secondary is 
\begin{equation}
 v_{\rm rel} = \sqrt{v_{\rm sec}^{\rm 2}+v_{\rm exp}^{\rm 2}} .
\label{eq:vrel}
\end{equation}
As $M_2/\msun \simeq R_2/\rsun$ the ratio of accretion radius
(\ref{eq:racc}) to stellar radius $R_2$ is 
\begin{equation}
   \frac{R_{\rm A}}{R_2} \simeq \left( \frac{360 {\rm km/s}}{v}\right)^{2}
\label{eq:rrapprox}
\end{equation}
(where $v$ is the quadratic mean of $v_{\rm sec}$, $v_{\rm exp}$ and
$c_{\rm S}$), suggesting that $R_{\rm A} < R_2$ for most cases.

Setting $c_{\rm drag} = 2$ and $R_{\rm A} = R_2$ we obtain for the
loss of angular momentum due to friction from the tangential
component of the drag force (\ref{eq:fdrag}) 
\begin{equation} 
 \dot{J}_{\rm FAML} = - |\vec{a} \times \vec{F}_{\rm drag}| = - a \: \pi \: 
	R_{\rm 2}^{\rm 2} \: \rho \: v_{\rm sec} \: v_{\rm rel} 
.
\label{eq:jdotfaml}
\end{equation}
Using the continuity equation 
\begin{equation}
 - \dot{M}_{\rm 1} = 4 \pi \: r^{\rm 2} \rho \: v_{\rm exp}
\label{eq:m1dot}
\end{equation}
to eliminate $\rho$, (\ref{eq:vsec}) and (\ref{eq-2.1.3})  
then give
\begin{equation} 
 \dot{J}_{\rm FAML} = \frac{1}{4} \: \left[f_{\rm 2}(q)\right]^{\rm 2} \: 
	\frac{v_{\rm rel}}{v_{\rm exp}} \: \dot{M}_{\rm 1} \: 
	\sqrt{G\:M\:a}
.
\label{eq:jdotfaml2}
\end{equation}

If most of the mass and angular momentum can be assumed to be lost at
roughly constant rate and velocity (Livio et al.\ 1991) 
we finally obtain the total FAML, integrated over the outburst, 
\begin{equation} 
 \Delta J_{\rm FAML} =
        - J \: \frac{\mej}{M_{\rm 1}} \: A(q) \: B(K_{\rm 1}) 
 \label{eq:deltafaml}
,
\end{equation}
where 
\begin{equation}
 A(q)=\frac{1+q}{4}\left[f_{\rm 2}(q)\right]^{\rm 2} 
%	\simeq 3^{\rm -\frac{8}{3}}\left(1+q\right)^{\rm \frac{1}{3}}
,
\label{eq:aofq}
\end{equation}
\begin{equation}
 B(K_{\rm 1})= \frac{\sqrt{1+K_{\rm 1}^{\rm 2}}}{K_{\rm 1}}=\frac{v_{\rm rel}}{v_{\rm exp}} 
 \label{eq:k1depend}
,
\end{equation}
and
\begin{equation}
 K_{\rm 1}=\frac{v_{\rm exp}}{v_{\rm sec}}
.
\label{eq:k1def}
\end{equation}

Comparison with (\ref{eq-2.3.1}) then shows that
\begin{equation}
 \nu_{\rm FAML} = \frac{1+q}{q}A(q)B(K_{\rm 1})
 \label{eq:nufaml}
,
\end{equation}
i.e.\ the mean specific a.m.\ $j_{\rm ej}$ of the ejected material is 
\begin{equation}
 j_{\rm ej} = ( \frac{1}{q} + \nu_{\rm FAML} ) \: \frac{J}{M}
 \label{eq:jej}
.
\end{equation}

\begin{figure}
\centerline{\plotone{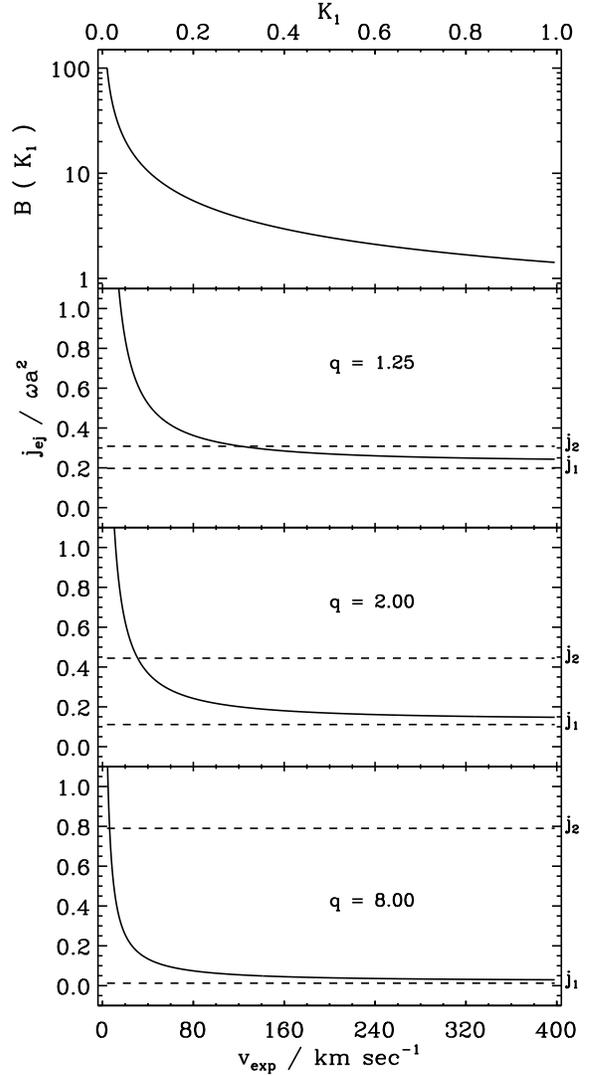}}
%\centerline{\refplotone{j_valid_new.eps}}
\caption[ ] 
{Top panel: the function $B(K_1)$ defined in (\protect\ref{eq:k1depend}).
Lower panels:
the specific angular momentum $j_{\rm ej}$ of the ejected matter
as a function of $K_{1}$ (or $v_{\rm exp}$, obtained by
assuming a constant $v_{\rm sec} = 400$ km/sec), measuring the   
strength of FAML, for different mass ratios $q=M_{\rm 1}/M_{\rm 2}$,
according to (\protect\ref{eq:jej}).
The dashed lines mark the specific a.m.\ $j_{\rm 1}$ and $j_{\rm 2}$ 
of the WD and the secondary, respectively. 
}
\label{fig:j_valid}
\end{figure} 

The functional dependence on $K_{\rm 1}$ in Eq.~(\ref{eq:k1depend}),
shown in Fig.~\ref{fig:j_valid} (top panel),
leads to the unphysical limit $-\Delta J_{\rm FAML}
\rightarrow \infty$ for a very slow envelope expansion $v_{\rm exp}
\rightarrow 0$. 
This is mainly due to the neglect of the envelope spin--up which
would effectively reduce the relative velocity (Livio et al.\ 1991). 

The secondary cannot spin the envelope up to speeds faster than
corotation, so $j_{\rm ej}$ is certainly smaller than 
$j_1 + a^2 \omega$ (where $\omega$ is the binary's angular velocity). 
If only a certain fraction of the envelope is spun--up to corotation
this upper limit is correspondingly smaller, e.g.\ 
\begin{equation}
 j_{\rm max} = j_1 + (R_2/a) a^2 \omega 
 \label{eq:jmax}
\end{equation}
if only the torus traced by the secondary's orbital motion corotates.

Equation~(\ref{eq:nufaml}) specifies our model of FAML in terms of a
single parameter, $K_1$ (or $v_{\rm exp}$) and determines the value 
of $\zr^F$ according to Eq.~(\ref{eq-2.3.7}). 
We use (\ref{eq:jmax}) as a physical upper limit on the effect of FAML.

\subsection{The effect of individual nova outbursts}
%++++++++++++++++++++++++++++++++++++++++++++++++++++++++++++++++++++++++++++++++++++++

\begin{figure}
\centerline{\plotone{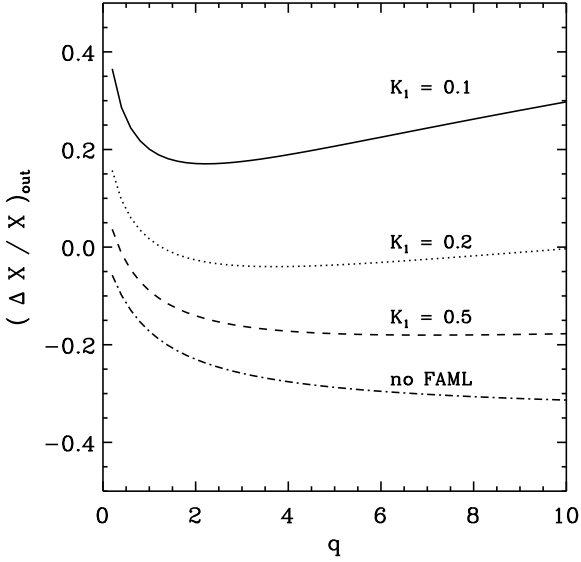}}
\caption[ ] 
{Outburst amplitude of the mass transfer rate according to
(\protect\ref{eq:dmm}) for various strengths of FAML (as
labelled) and $\Delta M_{\rm ej} = 10^{\rm -4}\:M_{\odot}$, $M_{\rm
1} = 1.2\: M_{\odot}$, $H_{\rm p} / R_{\rm R} = 1.4\:10^{\rm -4}$. }
\label{fig:m_jumps}
\end{figure}

To illustrate the effect of FAML in the parametrization
(\ref{eq:nufaml}) we consider the resulting relative
change $\Delta x/x = (x({\rm post})-x({\rm pre}))/x$ of a system 
parameter $x$ as a result of the nova outburst. 
For the orbital distance we obtain, similar to
(\ref{eq-2.3.2}),  
\begin{eqnarray} 
 \label{eq:daa}
 \lefteqn{
 \left(\frac{\Delta a}{a}\right)_{\rm out} =
	\frac{\mej}{M_{\rm 1}} \left( 
	\frac{q}{1+q} - 2\:A(q)\:B(K_{\rm 1}) \right) \simeq 
 } \\ \nonumber
	\lefteqn{ \quad \quad \quad
        10^{-4} \left( 1 - 0.2\: B(K_{\rm 1}) \right)
	\frac{\mej/10^{-4}\msun}{m_1} \: ,
}
\end{eqnarray}
which translates into 
\begin{eqnarray}
 \label{eq:dpp}
 \lefteqn{
 \left( \frac{\Delta P}{P}\right)_{\rm out} =
	\frac{\mej}{M_{\rm 1}} 
	\left( 2 \frac{q}{1+q} - 3\:A(q)\:B(K_{\rm 1}) \right) \simeq 
 } 
\\ \nonumber
	\lefteqn{ \quad \quad \quad
        10^{-4} \left( 2 - 0.6\:B(K_{\rm 1}) \right)       
	\frac{ \mej/10^{-4}\msun }{m_1}  \: ,
}
\end{eqnarray}
for the orbital period $P$. The approximate expressions in
(\ref{eq:daa}) and (\ref{eq:dpp}) make use of the weak dependence
on $q$; $m_1$ is $M_1/\msun$. For a given ejection 
mass and white dwarf mass, $B(K_1)$ critically determines the
magnitude of the orbital change.  
From (\ref{eq-2.4.2}) and (\ref{eq-2.3.2})
the change of the mass transfer rate $X=-\dot{M}_{\rm 2}$ is
\begin{eqnarray}
 \label{eq:dmm}
 \lefteqn{ 
	 \left(\frac{\Delta X}{X}\right)_{\rm out} 
	\simeq -\frac{\Delta R_{\rm R}}{H_{\rm p}} =
 } \\ \nonumber \lefteqn{ \quad \quad \quad
	- \frac{\mej}{M_{\rm 1}} 
	\left( \frac{q}{1+q} - \beta_{\rm 2}(q) - 2\:A(q)\:B(K_{\rm 1}) \right) 
 	\frac{R_{\rm R}}{H_{\rm p}} 
 } \\ \nonumber \lefteqn{ \quad \quad \quad
        \simeq - \left( 1 - 0.2\:B(K_{\rm 1}) \right)
	\frac{\mej/10^{-4}\msun}{m_1} \: , 
 }	
\end{eqnarray}
where we used $\epsilon \simeq 10^{-4}$ to obtain the last
line. Figure~\ref{fig:m_jumps} illustrates Eq.~(\ref{eq:dmm}).

\begin{figure}
\centerline{\plotone{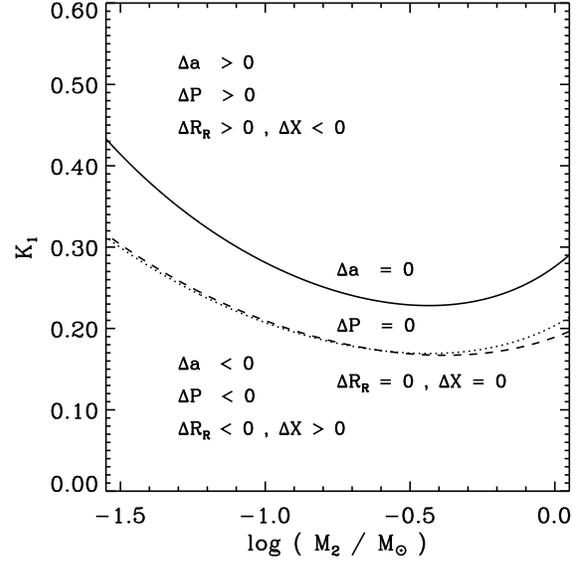}}
\caption[ ] 
{Critical values of $K_{\rm 1}$ where outburst amplitudes resulting from
a nova event just vanish, plotted as function of $M_{\rm 2}$
calculated with $M_{\rm 1}=1.2\:M_{\rm \odot}$.
Shown are the lines for orbital separation $a$ (full), orbital period
$P$ (dotted) and Roche radius $R_{\rm R}$ (dashed, same as for
mass transfer rate $-\dot{M}_{\rm 2}$)}
\label{fig:k1crit}
\end{figure}

Depending on the strength of FAML and the mass ratio the changes in
system parameters can be either positive or negative. The critical
$K_{\rm 1}(q)$ where the amplitudes vanish are shown in
Fig.~\ref{fig:k1crit}. The critical lines for $P$ and $X$ almost
coincide and separate the two different outburst types, 
those which increase the mass transfer rate {\em and} decrease the
orbital period, and those which decrease the transfer rate {\em and}
increase the period.

\subsection{The effect on the long--term continuous wind average}
%++++++++++++++++++++++++++++++++++++++++++++++++++++++++++++++++++++++++++++++++++++++

To consider the effect FAML has on the continuous wind average
evolution we first note that FAML is a
consequential angular momentum loss (CAML; cf.\ King \& Kolb 1995)
as $\dot J_{\rm FAML}$ is proportional to the mass transfer rate, see 
(\ref{eq:deltafaml}) with $\mej = X_{\rm FAML} t_{\rm rec}$ (where $X$
denotes the 
continuous wind average mass transfer rate). As such
FAML amplifies the transfer rate which would be driven by systemic angular
momentum losses alone in the absence of FAML. 

\begin{figure}
\centerline{\plotone{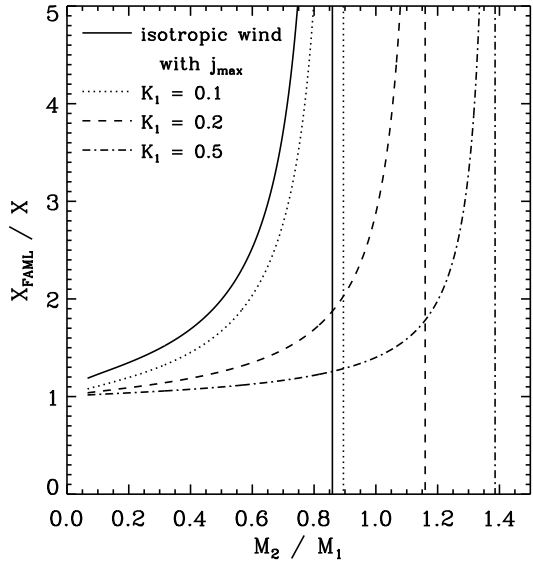}}
\caption[ ]{FAML amplification factor $X_{\rm FAML}/X$, defined in 
(\protect\ref{eq:mdotratio}), as a function of $M_2/M_1=1/q$, for
various values of $K_1$. The vertical lines indicate (thermal)
instability. 
The full line is an approximate global upper limit, independent of
$K_1$.  
}
\label{fig:factor_fix}
\end{figure}

An analytic estimate of this amplification factor, the ratio $X_{\rm
FAML}/X$ of the mass transfer rate with and without FAML,
can be obtained if we assume that the system follows the corresponding
uniform evolutionary track found by Stehle et al.\ \cite{stehle:etal} also in the
case with FAML. This is a reasonable assumption as long as
the amplification is moderate and the resulting mass transfer
timescale longer than the thermal time of the secondary. The radius
reaction of the secondary along its track is given by (\ref{eq-2.4.8}). 
Hence using $\dot R_{\rm R}/R_{\rm R}=\dot R_2/R_2$ and 
(\ref{eq-2.1.7}) we have 
\begin{equation} 
 \frac{X_{\rm FAML}}{X} \simeq 
      	\frac{\left(\zeta_{\rm e}-\zr\right)}
	    {\left(\zeta_{\rm e}-\zr^F \right)}  \simeq
	\left( 1 - \frac{2\:\frac{1}{q}\:A(q)\:B(K_{\rm 1})}
	{\zeta_{\rm e}-\zr} \right)^{\rm -1} \: .
 \label{eq:mdotratio}
\end{equation}
Here $\zeta_{\rm e} \simeq 0.85 \simeq$~const.\, and $\zr$ is from 
(\ref{eq-2.1.14}) with $\eta = 0$, $\nu=1/q$.
We emphasize that the ratio $X_{\rm FAML}/X$ does not depend on the
mass $\mej$ ejected per outburst, rather only on $K_1$ and the
mass ratio $q$. Fig.~\ref{fig:factor_fix} shows $X_{\rm FAML}/X$
as function of $1/q$ for various values of $K_1$. Obviously FAML
amplification is always small for large $q$ (small $M_2/M_1$). For a
given $K_1$ it increases with decreasing $q$ and formally goes to
infinity where the system approaches thermal instability ($\zeta_{\rm e}
= \zr^F$). 

The full line in Fig.~\ref{fig:factor_fix} indicates 
the upper limit for $X_{\rm FAML}/X$ 
if the specific angular momentum the ejected envelope can carry is
limited by $j_{\rm ej} \la j_1 + (R_2/a) a^2 \omega$ (see Sect.~3.1).

\section{Numerical examples} 
%***********************************************************

Here we apply the FAML description introduced in the previous section
and test the analytical considerations of Sect.~2 and 3 explicitly 
with numerical sequences of the secular evolution of CVs.

The above FAML model contains $K_1$,
$\mig$ and $\mej$ as free parameters. Unfortunately both
observations and theoretical models describing the outburst itself
yield often contradicting and inconsistent estimates for these 
parameters. Therefore we
restrict the following investigation to a simple parameter study to
obtain  
a systematic picture of the effect of FAML with a given strength on 
the secular evolution.
In particular we
present calculations with constant $\mig$ ($=\mej$) and $K_1$,
for various values of $\mig$ and $K_1$.

\subsection{Computational technique}
%+++++++++++++++++++++++++++++++++++++++++++++++++++++++++++

To model the binary evolution numerically we describe the secondary
star by either full stellar models  using Mazzitelli's 
stellar evolution code (e.g.\ Mazzitelli 1989), or by a simplified 
bipolytrope structure using the generalized bipolytrope code (Kolb \&
Ritter 1992). Both codes have been modified to include
FAML and to resolve individual nova outbursts. 

The bipolytrope code is about 4 orders of magnitude faster than the full
code and allows one to compute the secular evolution for several Gyr 
with every single nova outburst fully resolved. Several sequences
calculated with the full code serve as an independent check of the
simplified description. Despite a careful calibration of the bipolytrope
model to full stellar models, allowing a quantitative match of the
results to usually better than 10\%,  there are well--known
limitations of the simplified description (see Kolb \& Ritter 1992),
e.g.\ the increasing deviation of $\za$ from the values of actual
stars for $M_{\rm 2} \ga 
0.6\:M_{\odot}$. But these effects are negligible for the purpose of
this paper.

In all the examples shown below we compute magnetic braking 
according to Verbunt \& Zwaan \cite{verbunt:zwaan} with the
calibration parameter set to unity.
 
We have performed calculations with the following two modes:
\begin{enumerate}
\item {\sc fully resolved:} 
	Mass transfer is conservative until the mass	
	accreted on the WD exceeds the ignition mass. In the time step
	immediately thereafter the ejecta mass and the angular
	momentum (\ref{eq:deltafaml}) is removed 
	from the system. Thus fully resolved evolutions are
	discontinuous and show repeating nova cycles.
\item {\sc continuous wind average:} 
	Mass loss from the system is continuous at a rate equal to the
	transfer rate, carrying the 
	{\em increased} specific angular momentum according to
	(\ref{eq:jej}); hence the secular evolution is continuous. 
\end{enumerate}

\subsection{Detailed examples for individual outbursts}
%++++++++++++++++++++++++++++++++++++++++++++++++++++++++++++++++++++++++++++++++++++++

\begin{figure}
\centerline{\plotone{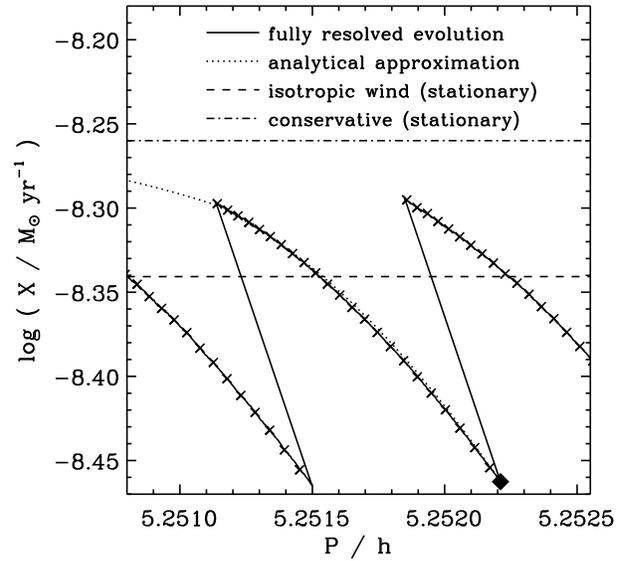}}
\caption[ ] 
{Section of a fully resolved evolutionary calculation with nova 
outbursts after
$\mig = \mej = 10^{\rm -4} M_{\odot}$ with
$M_{\rm 1}=1.2\:M_{\odot}$, $M_{\rm 2} \simeq 0.6\:M_{\odot}$ and 
weak FAML ($K_{\rm 1} = 0.5$). Crosses indicate time intervals of
1000 years, the filled diamond the post--outburst model from which
initial values for the 
analytical approximation (dotted line) were taken. 
The additional horizontal lines are explained in the text.} 
\label{fig:down}
\end{figure}

\begin{figure}
\centerline{\plotone{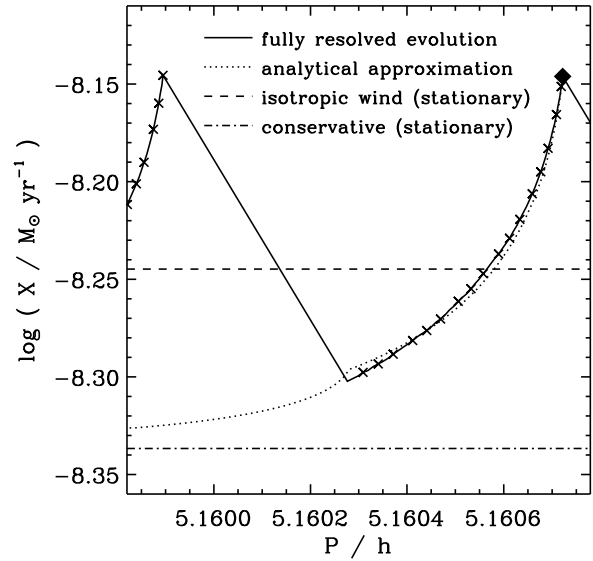}}
\caption[ ]{As Fig.\protect\ref{fig:down}, but with strong FAML
($K_{\rm 1}=0.1$).} 
\label{fig:up}
\end{figure}

We generated short stretches of fully--resolved evolutionary sequences 
with full stellar models by switching into the fully resolved mode 
in the middle of a continuous wind average calculation for a typical
CV above the 
period gap. The mode switch was made from a model
well--established on the uniform evolutionary track characterized by
(\ref{eq-2.4.8}).

Fig.~\ref{fig:down} shows for the case of weak FAML with $K_{\rm 1} =
0.5$ a 'sawtooth' (or 'shark--fin') like modulation of the mass transfer
rate over the orbital period (full line). The system evolves from
right to left; the crosses mark intervals of 1000 yr. As expected
from Fig.~\ref{fig:k1crit} the outburst results in a {\em longer}
period and a {\em lower} mass transfer rate with this choice of 
parameters. 
In the case of strong FAML with $K_{\rm 1} = 0.1$
(Fig.~\ref{fig:up}) the outburst leads to a {\em shorter}
period and a {\em higher} mass transfer rate. 

The approximate time evolution according to
Eq.~(\ref{eq-2.2.1}), with ${\cal A}$ and ${\cal B}^c$ evaluated
from the full sequence at the post--outburst position, marked with a
filled diamond, is shown as a dotted curve and matches the full
sequence very well. 
The dash--dotted line in Figs.~\ref{fig:down} \& \ref{fig:up} indicates the
level of the 'local' stationary conservative mass transfer rate as
given by Eq.~(\ref{eq-2.1.15}) with $\zr = \zr^c$. 
Note that there is a difference between this value
and the mass transfer rate of a completely conservative evolution with
the same initial values (but no outbursts) as the system here does
{\em not} evolve conservatively on average but loses mass together
with angular momentum.  
The averaged stationary mass transfer rate which includes FAML with 
$\zr^F$ according to Eq.~(\ref{eq-2.3.7}) is shown as 
a dashed line. This value is identical to the mass transfer rate
calculated in the averaged mode and to the time average of the full
evolution. 
The system approaches the conservative value between outbursts and oscillates
around the average transfer rate. 

Note that if, beginning with negligible FAML, the strength of FAML is 
continuously increased, the outburst amplitudes will decrease until 
the stationary (local) conservative transfer rate and the continuous 
wind average (which grows with FAML) are equal.
Further increase of FAML will now lead to growing amplitudes, but 
with opposite signs.

\subsection{Long--term evolution with FAML}
%++++++++++++++++++++++++++++++++++++++++++++++++++++++++++++++++++++++++++++++++++++++

\begin{figure}
\centerline{\plotone{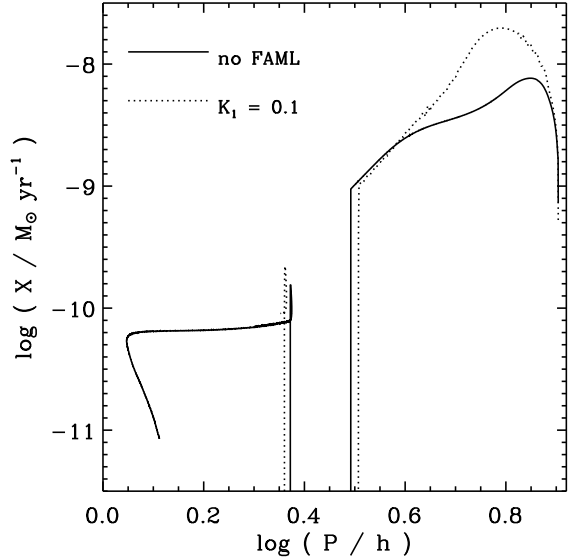}}
\caption[ ]{Averaged mass transer rate $X$ 
versus orbital period $P$ for a CV with $M_{\rm 1}=1.2\:M_{\odot}$ and
$M_{\rm 2}=1.0\:M_{\odot}$ at turn--on without FAML (full 
line) and with strong FAML ($K_{\rm 1}=0.1$, corresponding to 
$v_{\rm exp} \simeq 40$ km/sec; dotted line).
Computed with full stellar models.}
\label{fig:mdotpfaml}
\end{figure}

Fig.~\ref{fig:mdotpfaml} compares the continuous wind average
evolution for a reference system ($M_{\rm 1}=1.2\:M_{\odot}$; 
$M_{\rm 2}=1.0\:M_{\odot}$ at turn--on; $\eta=0$) with strong 
FAML and without FAML, computed with full stellar models. 
As expected from Fig.~\ref{fig:factor_fix} the mass transfer rate 
with FAML is significantly larger than in the case without only
at long periods, where
the system is close to (but not beyond) thermal instability. 
Due to the higher mass transfer rate the secondary in the FAML sequence
is driven further out of thermal equlibrium and therfore larger at the
upper edge of the period gap. At the same time the secondary's mass 
is smaller upon entering the detached phase.
Both effects, well--known from systematic studies of CV evolution
(e.g.\ Kolb \& Ritter 1992),
cause a wider period gap by both increasing the period at the upper
edge and decreasing the period at the lower edge of the gap.

\begin{figure*}
\centerline{\plottwo{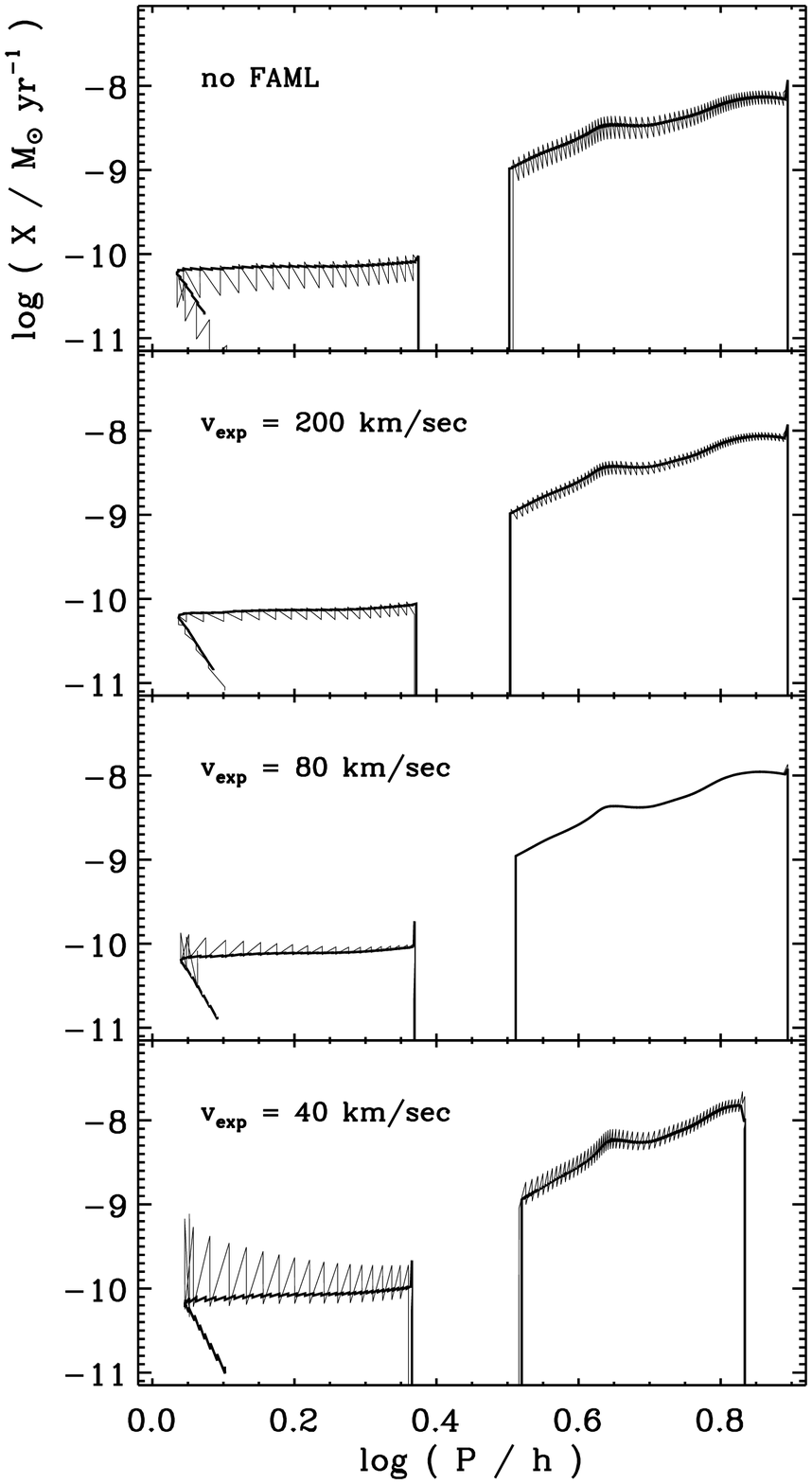}{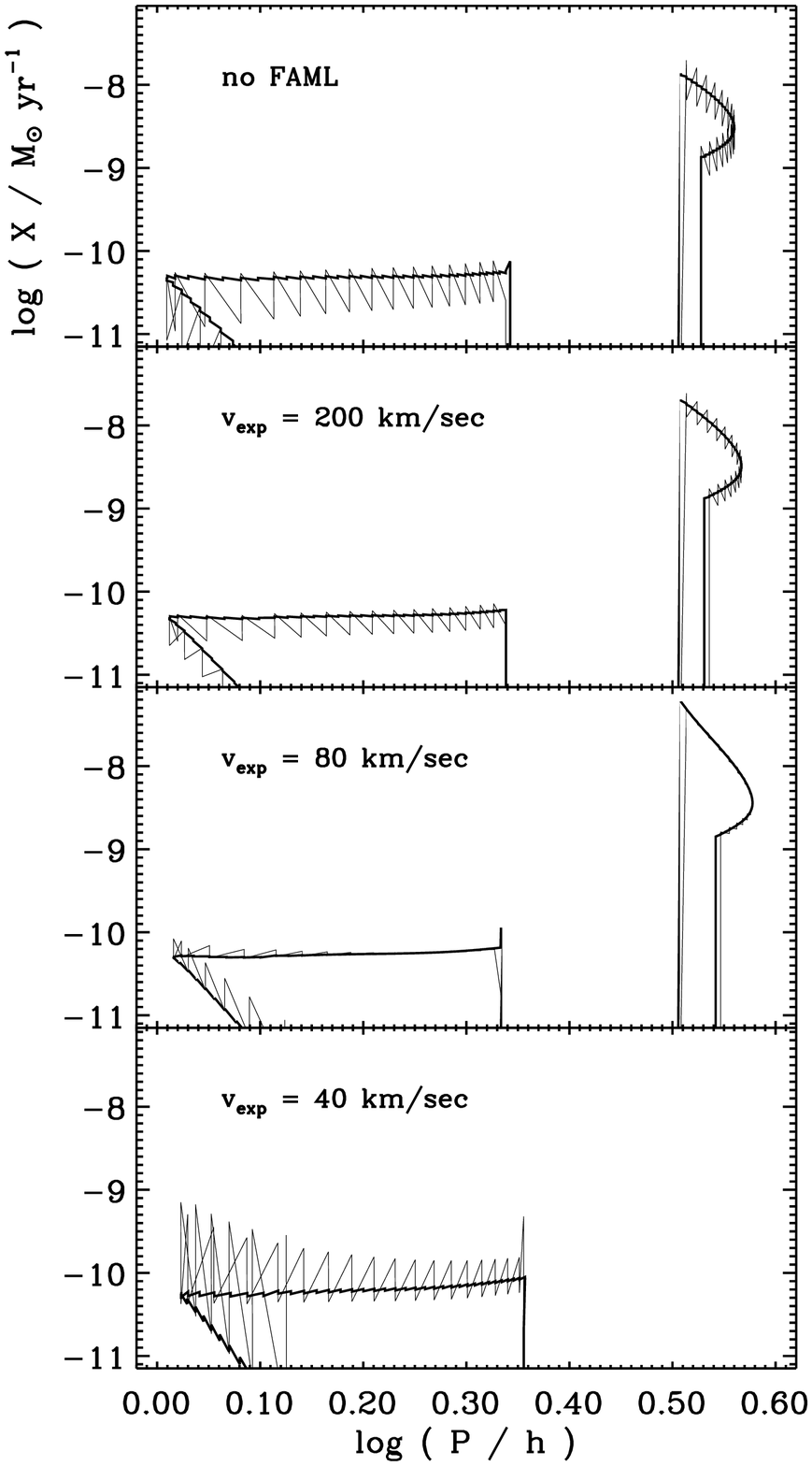}}
%\centerline{\refplottwo{satz1.eps}{satz2.eps}}
\caption[ ] 
{Set of fully resolved secular evolutions
of a CV with $M_{\rm 1}=1.2\:M_{\odot}$ (left) and $M_{\rm
1}=0.65\:M_{\odot}$ (right), computed
with the bipolytrope code. 
From top to bottom the strength of FAML increases as labelled. 
The evolutionary tracks connect the location immediately before and
after an outburst, {\em but not for subsequent outbursts}!
Rather only every every $1000^{\rm th}$
outburst (thick line, corresponding to $\Delta M_{\rm ig}
= 5\:10^{\rm -6} M_{\odot}$), or every $100^{\rm th}$ outburst
(thin line, corresponding to $\Delta M_{\rm ig} = 10^{\rm -4}
M_{\odot}$) is shown.
The initial secondary mass at turn--on was $M_{\rm 2}=1.0\:M_{\odot}$ for the top
three sequences on the left, and $0.4\:M_{\odot}$ on the right. For
the sequences in the lowest panel the initial secondary mass was 
$0.85\:M_{\odot}$ and $0.25\:M_{\odot}$, respectively.}
\label{fig:satz1}
\end{figure*}

The analytical considerations of Sect.~3.3 are confirmed by a
parameter study performed with the bipolytrope code.
The evolutionary sequences with all outbursts resolved depicted in
Fig.~\ref{fig:satz1} summarize the results. 
The reference system is shown in the left column, a less massive 
system in the right column. 
The strength of FAML increases from top to bottom, with consequences for
the outburst amplitudes as discussed in Sect.~3.2 and in the previous
subsection. Note that the 'sawtooths' here {\em do not represent} single 
outburst cycles, but arise because we connect the pre-- and post--outburst
values for every n$^{\rm th}$ ($n=1000$ or $100$) outburst only.
The advantage of this display method is that both 
magnitude and sign of the outburst amplitude
(the vertical flanks) can be followed through the evolution. 
In reality the system undergoes many individual cycles like those 
in Figs.~\ref{fig:down} \& \ref{fig:up} between subsequent 
outbursts depicted in Fig.~\ref{fig:satz1}.
The two evolutionary tracks plotted in each panel differ only in 
the ignition mass $\mig$ (here $= \mej$). The thick line
corresponds to $\mig = 5\times10^{\rm -6}\:M_{\odot}$, the thin 
line to $\mig = 10^{\rm-4}\:M_{\odot}$. In the scale used the low
ignition mass  
sequences are practically indistinguishable from the continuous wind
average evolution 
(remember that both oscillate around the {\em same average value}, 
cf.\ Sect.~\ref{sec-4.5}). 
Inspection of Fig.~\ref{fig:m_jumps} shows that the growth of
amplitudes towards 
shortest periods is mainly due to the increase of $q$. Only in the
high FAML (low $v_{\rm exp}$) case, i.e.~in the very steep region of
Eq.~(\ref{eq:k1depend}), there is an additional contribution from the 
increasing orbital velocity (hence decreasing $K_{\rm 1}$) at short $P$.

\subsection{FAML--induced instability}
%++++++++++++++++++++++++++++++++++++++++++++++++++++++++++++++++++++++++++++++++++++++

To illustrate and test the analytical mass transfer stability
considerations of 
Sects.~2.3 and 2.4 explicitly we consider the secular evolution of
a system which runs into an instability, i.e.\ violates the formal
stability criterion (\ref{eq-2.4.1}) at some point of the evolution.  

\begin{figure}
\centerline{\plotone{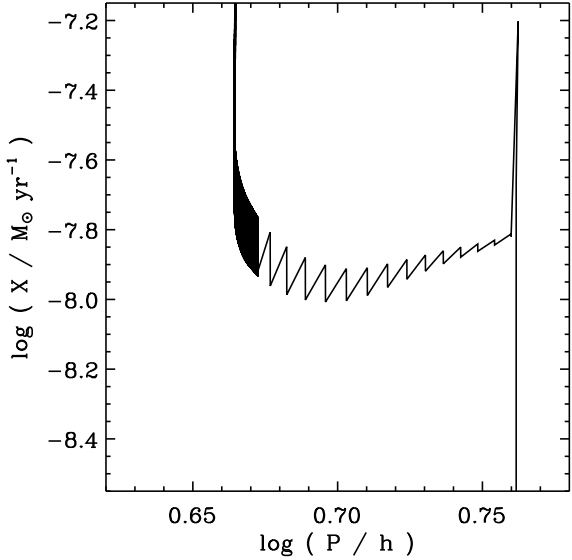}}
\caption[ ] {Fully resolved secular evolution of a CV with 
$M_{\rm 1}=0.8\:M_{\odot}$ and $M_{\rm 2}=0.7\:M_{\odot}$ at turn--on, 
computed with the bipolytrope code including FAML based on (\ref{eq:linvexp}) 
and $\mig = \mej = 10^{-4} \msun$. 
For $\log ( P/{\rm h} ) > 0.67$ every $100^{\rm th}$ outburst is shown,
for $\log ( P/{\rm h} ) < 0.67$ every outburst.
%%% (in fact every $20^{\rm th}$ model calculated). 
%The output density was increased in the 
%final phase until the calculation terminated shortly after 
%--> lieber Klaus, das ist zwar schoen kompakt und kurz, aber
%verstehen tut's keiner. Sag doch einfach praeziese was gemacht wird:
%jeder x-te Ausbruch gezeigt fuer P>.., fuer P< .. jeder y-te.
%the runaway began.}
}
\label{fig:runaway}
\end{figure}

\begin{figure}
\centerline{\plotone{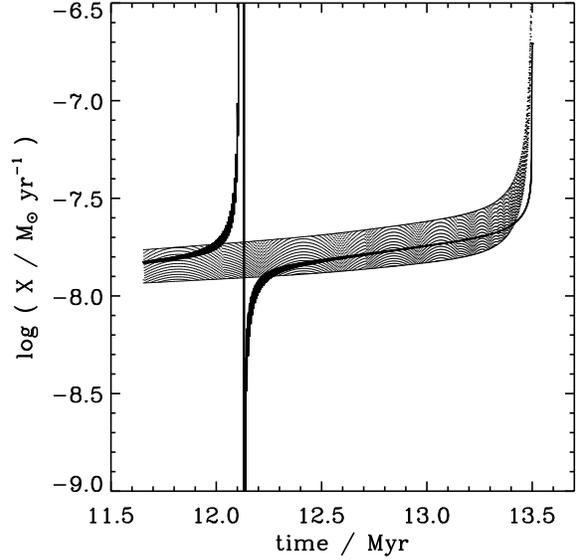}}
\caption[ ] {Evolution of the mass transfer rate $X$ with time during the 
final phase (approximately 2 million years) before the runaway (dotted,
filling the grey--shaded area). 
%
% This looks black to me on the Figure!
%
Overplotted in full linestyle is the stationary value 
$X_s^F = {\cal A} / {\cal B}^F$ for a continuous FAML-increased mass loss,
indicating an instability at $\simeq 12.1\:{\rm Myr}$.}
\label{fig:runzoom}
\end{figure}

To establish this situation we artificially increase the strength of
FAML along a fully resolved evolutionary sequence (computed with the
bipoltrope code, with constant $\mig = \mej = 10^{-4} \msun$), 
starting from a small value at the onset of mass
transfer. As a consequence $\zr^F$ becomes larger than $\za$ at some
intermediate secondary mass. In particular, we adopt
\begin{equation}
  \frac{v_{\rm exp}}{{\rm km/s}} = 180 \frac{M_2}{\msun} - 50  
\label{eq:linvexp}
\end{equation}
as the functional form of the envelope expansion velocity.
($M_2$ will be large enough to avoid negative $v_{\rm exp}$). 
The corresponding $\zr^F$ with $\nu_{\rm FAML}$ from 
(\ref{eq:nufaml}) becomes larger than $\za$ at $M_2 \simeq
0.55\:\msun$, and the mass transfer rate indeed begins to 
grow without limit at about this mass well above the period gap
(Fig.~\ref{fig:runaway}). 
A closer look at the final phase 
immediately before the runaway (Fig.~\ref{fig:runzoom}) shows that 
the system evolves beyond the formally unstable point apparently
unaffected and enters the runaway mass transfer phase only $\simeq
10^6 {\rm yr}$  
later. In fact this is just as we would expect if the evolution 
were replaced by the corresponding continuous wind average.
In this case (\ref{eq-2.1.9}) with ${\cal B} = {\cal B}^F$ describes the 
variation of the (average) mass transfer rate $X$.
The characteristic timescale on which $X$ changes is 
$X/\dot X = 1/({\cal A} - {\cal B}^F X)$, which becomes very large
when ${\cal B}^F \simeq 0$, i.e.\ around $\za = \zr^F$ at the
instability point; note that ${\cal A}$ is also small at this
point, see (\ref{eq-2.4.9}).
%Hence the mass transfer rate does not grow immediately. 
Proceeding further increases 
the strength of FAML even more and makes ${\cal B}^F$
more negative, hence the timescale $|X/\dot X|$ progressively
shorter. 
The time delay until the runaway finally begins is 
essentially determined by the crossing angle between the functions
$\za(M_2)$ and $\zr^F(M_2)$ 
%in Fig.~\ref{fig:zetarun} 
and the value of $X$ at this point.

%\begin{figure}
%\centerline{\plotone{zetarun_new.eps}}
%\caption[ ] {
%Comparison of both relevant mass--radius exponents for 
%$M_1 = 0.8\:\msun$ and FAML according to (\ref{eq:linvexp}). Their 
%intersecton marks the occurence of dynamical instability.}
%\label{fig:zetarun}
%\end{figure}

\begin{figure}
\centerline{\plotone{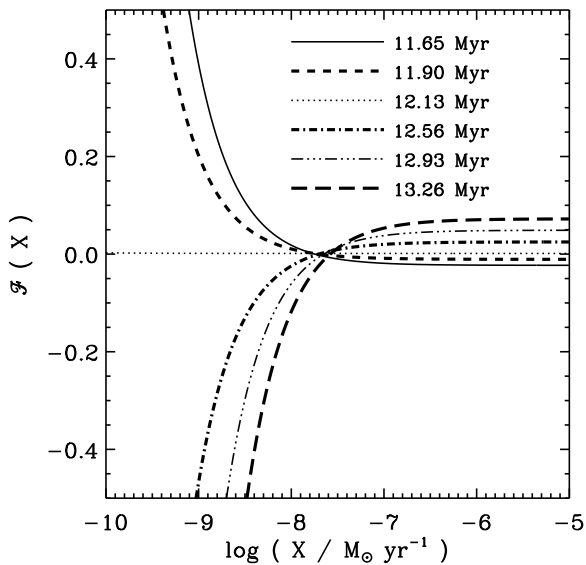}}
\caption[ ] {Growth function ${\cal F}$ (cf.~(\ref{eq-2.4.6}) and 
Fig.~\ref{fig:fhat}) calculated with data taken from the evolution
shown in the previous figures for different times 
%before and after
%the crossing of the instability 
as labelled.}
\label{fig:frun}
\end{figure}

Figure~\ref{fig:frun} shows the growth function ${\cal F}$ defined in
(\ref{eq-2.4.5}) for selected models immediately before the runaway
for the evolutionary sequence depicted in Fig.~\ref{fig:runzoom}. 
The different curves are computed with data from the run
taken at the times as labelled (cf.\ Fig.~\ref{fig:runzoom}).
As discussed in Sect.~\ref{sec-2.4} the change of ${\cal F}$ 
from a stable curve with negative slope to an unstable curve with
positive slope occurs when $\za = \zr^F$.

The sequence shown in Figs~\ref{fig:runaway}-\ref{fig:frun} is 
typical of the behaviour of evolutionary sequences with 
nova cycles and very strong FAML. It convincingly demonstrates 
that the FAML--amplified istropic wind average properly describes 
the effect of a sequence of nova outbursts, even in the most 
extreme case when the system approaches a FAML--induced instability. 
A more systematic investigation of the significance of this 
instability for the secular evolution of CVs will be published 
elsewhere (Kolb et al., in preparation).

\subsection{Evolution with FAML parameters from theoretical nova models}
\label{sec-4.5}
%++++++++++++++++++++++++++++++++++++++++++++++++++++++++++++++++++++++++

The FAML parameters $K_1$, $\mej$ and $\mig$ are 
certainly not constant along the evolution but depend on the actual
state of the outbursting system. The three governing parameters
which crucially determine the outburst characteristics of 
thermonuclear runaway (TNR) models for classical novae are the WD
mass $M_1$, the (mean) accretion rate $X$ and the WD temperature $T_1$ 
(e.g.\ Shara 1989). Both the mass and the temperature of
the WD change only slowly during the binary evolution, thus the dominant
dependence of $v_{\rm exp}$ is from the (mean) mass transfer rate. 
As FAML itself drives the mass transfer this could lead to an
interesting feedback and possibly self--amplification in the system, a
question considered in a separate paper (Kolb et al., in
preparation). 

Although published TNR models mostly give values for $\mig$ and
estimates for $\mej$, few or no data are available on $K_1$. 
Unfortunately this is also true for the 
most complete and consistent set of published nova models (Prialnik \&
Kovetz 1995) which we chose to use as input to the FAML
description derived in Sect.~\ref{sec-3.1}.

\begin{figure}
\centerline{\plotone{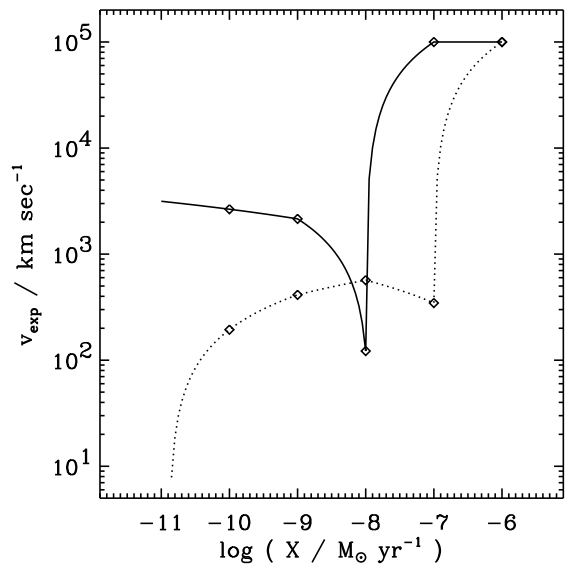}}
\caption[ ] 
{Averaged maximal expansion velocity of the ejecta taken from nova
models of Prialnik \& Kovetz \protect\cite{prialnik:kovetz}, with
$T_{\rm 1}=10^{\rm 7}$ K and $M_{\rm 1}=0.65\:M_{\odot}$ (full line) or
$M_{\rm 1}=1.25\:M_{\odot}$ (dotted line). Diamonds mark real
data points. A  maximum value of $10^{\rm 5}$ km/sec has
been inserted if no ejection occurred to mimic negligible FAML.}
\label{fig:vav}
\end{figure}

These authors tabulate only the average expansion velocity 
$v_{\rm av}$, the time average of the {\em maximum} velocity 
in the flow during the whole mass loss phase. 
In the absence of more relevant velocity data we simply
set $v_{\rm exp} = v_{\rm av}$, but we are aware that 
this choice represents an upper limit for $v_{\rm exp}$ 
(which has to be taken at the secondary's position). Two exemplary
relations $v_{\rm exp}(X)$ are shown in Fig.~\ref{fig:vav}, for a low
and a high mass WD, complemented by large values ($10^5$
km/sec, hence negligible FAML) when no envelope ejection was found in 
the hydrodynamical simulations. 

\begin{figure}
\centerline{\plotone{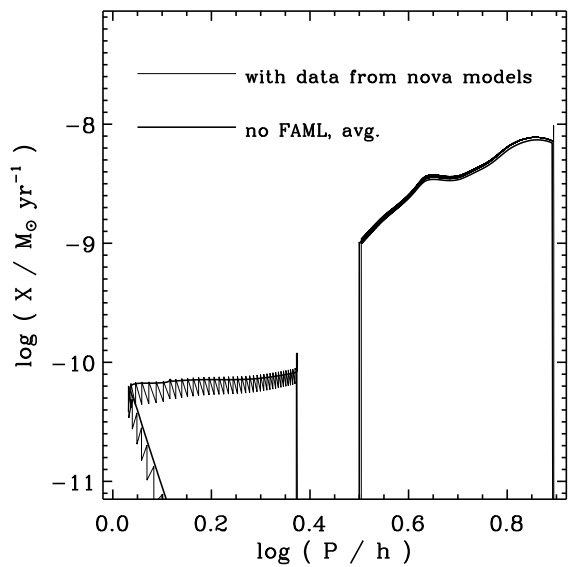}}
\caption[ ] 
{Secular evolution of a CV with fully resolved nova cycles and $M_{\rm
1}=1.2\:M_{\odot}$, $M_{\rm 2}=1.0\:M_{\odot}$ at turn--on (thin 
line), computed with the bipolytrope code. Ignition mass, ejection
mass and velocities are taken from 
Prialnik \& Kovetz \protect\cite{prialnik:kovetz}. Similar to
Fig.~\protect\ref{fig:satz1} only every $100^{\rm th}$ outburst is shown,
and straight lines connect post-- and pre--outburst points.
For comparison the continuous wind average evolution 
without FAML (thick line) is also shown. At the period minimum the 
WD mass has decreased to $M_1 \simeq 1.1\:\msun$.}
\label{fig:dina_high}
\end{figure}

\begin{figure}
\centerline{\plotone{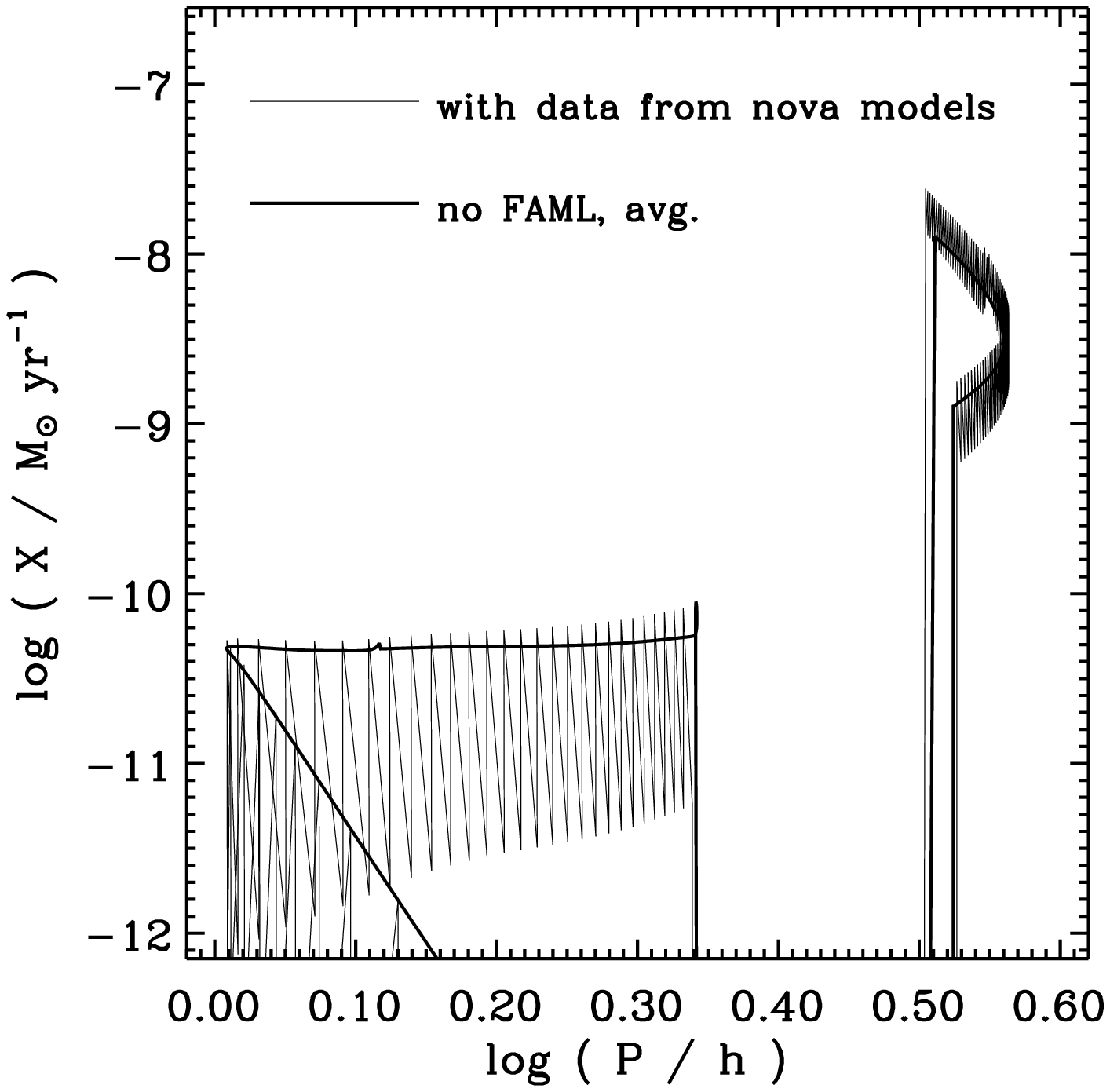}}
\caption[ ] 
{As Fig.~\protect\ref{fig:dina_high}, but with $M_{\rm 1} =
0.65\:M_{\odot}$, $M_{\rm 2.i} = 0.4\:M_{\odot}$, and every 
$20^{\rm th}$ outburst shown. At the period minimum the 
WD mass has decreased to $M_1 \simeq 0.635\:\msun$.}
\label{fig:dina_low}
\end{figure}

The evolutionary sequences obtained with $M_{\rm 1} = 1.2 \:
M_{\odot}$ (using the 
velocity data  
for nova models with a $1.25 \: M_{\odot}$ WD) and with $M_{\rm 1} = 0.65
\: M_{\odot}$ are 
shown in Figs.~\ref{fig:dina_high} \& \ref{fig:dina_low}. 
The changes 
of $M_{\rm 1}$ during the whole evolution are very small ($\eta \simeq
0$). As in Fig.~\ref{fig:satz1} the 'sawtooths' result 
from connecting only every ${\rm n}^{\rm th}$
outburst. For comparison  
a standard continuous wind average evolution without FAML but otherwise
identical parameters has been overlaid in thick linestyle. 
As expected from the high expansion velocities FAML has very little
effect on the global evolution. 
By choice we underestimated the strength of FAML in the sequences shown
above. But although the expansion velocity at the location
of the secondary is indeed smaller than $v_{\rm av}$ in the models by
Prialnik \& Kovetz \cite{prialnik:kovetz} 
they currently appear to be too large to cause a FAML effect with
significance for the long--term evolution of CVs 
(cf.\ also discussion in next section).

Perhaps the most interesting feature of the sequences presented in
this paragraph is the large outburst amplitude of the mass transfer
rate in low--mass WD ($0.65 \msun$)
systems below the period gap. The reason for this lies mainly in the fact 
that $\mej$ and $\mig$ to a first approximation scale like $R_1^4/M_1$ 
(Fujimoto 1982, $R_1$ is the WD radius), i.e.\ increase with decreasing 
WD mass (and to a lesser extent, increase with decreasing $X$). 
Moreover the slow but continuous decrease of the WD mass leads to a
further increase of the outburst amplitudes towards shorter orbital
period in this sequence (Fig.~\ref{fig:dina_low}). 
Note that this effect adds to the ones observed before 
(lowermost panels of Fig.~\ref{fig:satz1}) which were caused by the change
of $q$ due to decreasing $M_2$, and to a lesser extent to the slowly 
increasing orbital velocity $v_{\rm sec}$, mimicing increasing FAML strength.

\begin{figure}
\centerline{\plotone{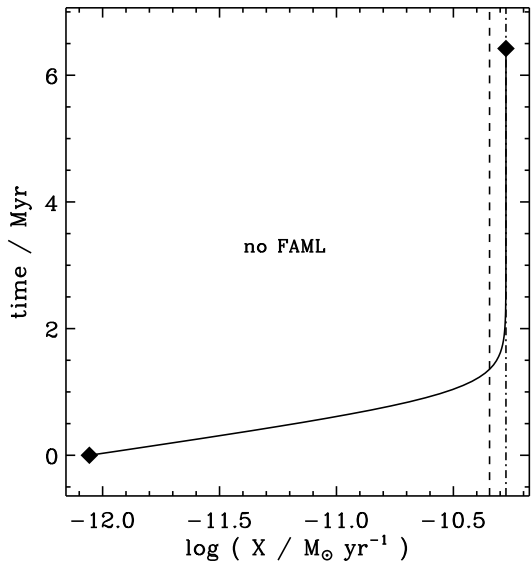}}
\caption[ ] 
{
A typical inter--outburst low--mass WD CV below the period gap 
($M_{1} = 0.6382\:\msun$, $M_{2} \simeq 0.1\:\msun$, $\mej = \mig =
2.9\:10^{-4} \msun$) according to nova parameters from Prialnik \&
Kovetz \cite{prialnik:kovetz}. Shown is the time spent with a mass transfer rate {\em
below} a value $X$, as a function of $X$ (in $\msun$/yr). Diamonds
indicate the immediate post- or pre--outburst state. The dashed line
indicates the continuous wind average mass transfer rate.
}
\label{fig:problow_t}
\end{figure}

Such amplitudes account for a non--negligible time interval with
a secular mean mass transfer rate 
significantly {\em below} the continuous wind average transfer rate,
i.e.\ below $\simeq 5 \times 10^{-11} M_\odot$/yr, the typical 
value for non--degenerate CVs below the period gap driven by gravitational
radiation alone. As an example Fig.~\ref{fig:problow_t} 
plots for a model at $P = 1.23$~h 
%--> please insert the correct value
from the sequence shown in Fig.~\ref{fig:dina_low}
the time spent with a mass transfer rate {\em
below} a value $X$, as a function of $X$. 
One nova cycle lasts $\simeq 6.5$~Myr, and  
the mass transfer rate remains for 0.5 Myr after the outburst 
below $10^{-11}\msun$/yr. Intrinsically, these CVs with low-mass and
moderate--mass WDs form the vast majority (e.g.\ de Kool 1992, 
Politano 1996). Thus, ignoring selection effects, we would
expect to observe 1 out of 10-20 short--period CVs with small sub--GR 
driven mass transfer rate. 

This seems to offer an alternative explanation for the low mass
transfer rate Sproats et al.\ \cite{sproats:etal} claim to find observationally in
so--called  
tremendous outburst amplitude dwarf novae (TOADs) which Howell et
al.\ \cite{howell:etal} interpret as post minimum period systems. As the secular
mean mass transfer rate is predicted to drop substantially ($\la
10^{-12} - 10^{-11} \msun$/yr) CVs which have evolved past the minimum
period would be even fainter than the above post--outburst CVs, and it
is unclear if they are detectable at all.

We note that at least 4 objects in the list of Sproats et al.\ 
\cite{sproats:etal} have been labelled as novae (cf.\ Duerbeck 1987): 
AL Com (orbital period $0.0566$~d,  
outburst suspected 1961), VY Aqr ($0.0635$~d, 1907), RZ Leo
($0.0708$~d, 1918) and SS LMi (no period known; 1980).
With exception
of SS LMi they are now firmly considered to be dwarf novae.  
This does not exclude a priori that the first recorded outburst
(e.g.\ in 1907 for VY Aqr) or even earlier ones were actually novae. 
Hence it is at least conceivable that the present low mass transfer
rate and the TOAD characteristics might be a consequence of this last
nova event. 
The three systems with known orbital period are located at the extreme
end of Fig.~4 of Sproats et al.\ \cite{sproats:etal}, showing large outburst
amplitude and low  
quiescence magnitude. Moreover, among the open square symbols
indicating DNe in their Fig.~3 they are also those with lowest period {\em
and} absolute magnitude in quiescence (mass transfer rate
$\ll 10^{-11} M_{\odot}$/yr). 
The fact that the claimed deviation from the secular mean magnitude is
shrinking with the time elapsed since the potential nova outburst
(1961 - 1918 - 1907) is probably just a coincidence.

We caution that
it is difficult to decide if the set of models by Prialnik \& Kovetz
\cite{prialnik:kovetz} properly describes classical nova outbursts on WDs in CVs
below the period gap. Observationally, only 4 out of 28 classical
novae with determined orbital period are below the gap (e.g.\ Ritter
\& Kolb 1995). This is in conflict with standard population models of
CVs if the Prialnik \& Kovetz \cite{prialnik:kovetz} ignition masses are used to
predict an observable period distribution for novae (Kolb 1995).

\section{Discussion and conclusions}
%%%%%%%%%%%%%%%%%%%%%%%%%%%%%%%%%%%%%

In this paper we considered the effects of nova outbursts on the
secular evolution of CVs. As a result of these outbursts
the secular mean mass transfer rate
and the orbital period are not continuous functions of time but change
essentially discontinuously with every nova outburst by an amount
proportional to the ejected envelope mass. 
In addition, energy and angular momentum can be removed from
the orbit due to dynamical friction of the secondary
orbiting in the expanding nova envelope. 

The discontinuous evolution with a given strength of 
frictional angular momentum loss (FAML) is usually
replaced by the corresponding continuous wind average evolution,
where the mass and angular momentum loss associated with a nova
outburst is assumed to be distributed over the inter--outburst 
time and to form an isotropic wind from the white dwarf. 
We showed analytically that the well--known mass transfer stability
criterion for the latter case can also be derived from a 
proper analysis of the real, discontinuous process, for an
arbitrary strength of FAML--amplification. 

We specified a quantitative model for FAML
within the framework of Bondi--Hoyle accretion following Livio
et al.\ \cite{livio:etal}. In this model 
the strength of FAML depends crucially on the
expansion velocity $v_{\rm exp}$ of the envelope at the location of
the secondary, being the stronger the smaller $v_{\rm exp}$ is.
We expect that the resulting simple one--parameter description 
properly describes the order of magnitude of the FAML effect and,
more importantly, the differential dependences on fundamental
binary parameters. 
Hence although it is a useful way to study the 
potential influence of FAML systematically, it certainly cannot
replace a detailed modelling of the frictional processes.

Calculations of the long--term evolution of CVs
verified the validity of the replacement of the discontinuous
sequence of nova cycles with the continuous wind average,
even for situations close to mass transfer instability, whatever the
strength of FAML. 
The mass transfer rate in the continuous wind average
evolution is FAML--amplified, i.e.\ by the factor (\ref{eq:mdotratio})
larger than  the transfer rate driven by systemic angular momentum 
losses alone. For a given CV this factor is determined by $K_1$ alone,
i.e.\ independent of the ejection mass. We emphasize that FAML 
only {\em amplifies} the transfer rate caused by systemic losses, it
does not {\em add} to them. Hence FAML is a particular example of 
consequential
angular momentum losses (CAML) investigated in detail by King \& Kolb
\cite{king:kolb}.  

In general, the FAML amplification factor turns out to be large only 
when the envelope expansion is very slow 
($K_1 \la 0.5$, i.e.\ $v_{\rm exp} \la 200$~km/s) 
{\em and} when the system is already close
to thermal mass transfer instability (Fig.\ref{fig:factor_fix}). This
latter condition means that 
e.g.\ for an evolution with strong FAML $K_1=0.1=$~const.\ the
averaged mass transfer rate is significantly affected only
at long orbital period, $P \ga 5$~h.

The magnitude and direction of the outburst
amplitudes of the mass transfer rate $X$ and the orbital period $P$
depend on both the ejection mass $\mej$ and 
the FAML parameter $K_1$.
For weak (or negligible) FAML the
outbursts are towards lower mass transfer rate and longer orbital
period, for strong FAML towards larger $X$ and shorter $P$. There is
an intermediate regime where the outburst amplitudes essentially
disappear. 

Theoretical models for nova outbursts generally find larger expansion
velocities (e.g.\ Prialnik 1986, Prialnik \& Kovetz 1995), 
typically $K_1 \ga 1$.
This is certainly true for the terminal velocities, but
in more recent models probably also for the crucial velocity 
at the secondary's location, i.e.\ closer to the WD. 
Kato \& Hachisu \cite{kato:hachisu} argue that in the
wind mass loss phase of a nova outburst the main acceleration (at
about the sonic point) takes place at a temperature where the opacity
has a maximum. Thus the introduction of the OPAL opacities had the
effect of moving this point closer to the WD. A comparison of the radial 
velocity profiles found by Kato \& Hachisu \cite{kato:hachisu} with those
obtained previously (e.g.~Prialnik 1986, Kato 1983) confirms this.
As a result, in the newer models the velocities at the location of the
secondary are already quite close to their
terminal values. This explains why Kato \& Hachisu
\cite{kato:hachisu} find only marginal effects from dynamical
friction.

This seems to suggest that the overall influence of FAML on
the long--term evolution of CVs is small. 
However, in view of the 
considerable simplifications of our FAML description and the
uncertainties of theoretical TNR models for nova outbursts,
it is worthwhile to investigate sytematically mass transfer stability
with FAML, and to consider the role of a feedback 
between outburst characteristics (hence FAML strength) and the mean
mass transfer rate prior to the outburst. 
We will study this in a forthcoming paper (Kolb et al., in 
preparation).  

To illustrate further the effect FAML might have on the secular
evolution of CVs we have calculated evolutionary sequences with
envelope expansion velocities, ignition and ejection masses taken from
the extended set of nova models by Prialnik \& Kovetz \cite{prialnik:kovetz}. As
expected, the continuous wind average evolution hardly differs
from the standard CV evolution without FAML. As a consequence of the
large ejection mass (several $\simeq 10^{-4} \msun$/yr)
the outburst amplitudes become very large (a factor $\ga 10$) below
the period gap for intermediate mass WDs ($M_1\simeq 0.6\msun$)
--- even more so as the outburst--induced decrease of the mass
transfer is largest if FAML vanishes.

Such systems have a mass transfer rate less than $10^{-11} \msun$/yr 
for $\simeq 0.5$~Myr after the outburst, and could account for
intrinsically faint CVs below the period gap. We speculated (Sect.~4.5)
if TOADs (e.g.\ Sproats et al.\ 1996) could represent such systems.

We finally note that the apparent scatter in observationally derived
values for the mass transfer rate of CVs with comparable orbital period,
well--known since the review of Patterson \cite{patterson}, is
unlikely to be due to outburst amplitudes.
First, we expect from Fig.~\ref{fig:dina_high} that these amplitudes are
small or negligible above the period gap, and second the systems spend
most of the time close to the continuous wind average mass transfer
rate (cf. Fig.~\ref{fig:problow_t}).  
A more promising explanation for this scatter assumes 
mass transfer cycles which could be irradiation--induced (e.g.\ King et
al.\ 1995, 1996).

\subparagraph{Acknowledgements.}
We thank D.~Prialnik and A.~Kovetz for giving access to nova model
parameters prior to publication, and M.~Livio and L.~Yungelson
for providing a copy of an unpublished manuscript on FAML. 
K.S.~would like to thank M.~Ruffert for many discussions on
Bondi--Hoyle accretion. We thank A.~King for improving the language of
the manuscript. K.S.~obtained partial financial support from the 
Swiss National Science Foundation. Theoretical astrophysics research 
at Leicester is supported by a PPARC rolling grant.

%===========================================================

% ----------------- End of Main Text -----------------------

% ----------------- Bibliography ---------------------------

% -------- Figures ('figure*' for full width)---------------

% ------------- Tables ('table*' for full width) -----------

% ---------------------- Appendix ---------------------------
\begin{appendix}

%%\section*{Appendix: Generalization of $\zr^F$}
\section{Generalization of $\zr^F$}
%++++++++++++++++++++++++++++++++++++++++++++++++++++++++++++

Starting from the fundamental Eq.~(\ref{eq-2.1.5}) one can derive a
more general expression for the 
relative change of $\rr$ during both the outburst and 
inter--outburst phase. Specifically we allow 
a small fraction $\alpha$ of the mass ejected during the 
outburst to accrete on to 
the secondary, i.e.\ $\Delta M_2 = \alpha \: \mej$. This gives 
\begin{eqnarray}
\label{eq:a1}
  {\left( \frac{\Delta \rr}{\rr} \right)}_{\rm out} 
	\lefteqn{ 
		= \frac{\mej}{M_2} 
	\left[ 	
		\frac{2-\beta_2}{q}
		- \alpha (\beta_2+2) -
	\right. }
\\ \nonumber
	\lefteqn{ \left. 
  		\frac{1-\alpha}{1+q}
		( 1 + \frac{2}{q} + 2\:\nu_{\rm FAML} )
	\right] }
\end{eqnarray}
instead of (\ref{eq-2.3.2}), which is obtained in the case $\alpha = 0$.

Additionally we allow a wind (or other) mass loss
during the inter--outburst accretion phase. 
Thus we use the general $\zr$ from (\ref{eq-2.1.14}) rather than the
conservative one to generalize (\ref{eq-2.3.4}). 
This results in two additional parameters, $\eta^w$ and $\nu^w$, 
describing the system's "wind" mass and corresponding angular momentum
loss during the inter--outburst phase. 
Hence we get (using $\gamma = (\eta^w+1)/\eta^w$)  
\begin{eqnarray}
\label{eq:a2}
  {\left( \frac{\Delta \rr}{\rr} \right)}_{\rm inter} 
	\lefteqn{ 
		=  2 \frac{\Delta J_{\rm sys}}{J} 
		+ \frac{\mig}{M_2} 
	\left[ 
		\frac{\beta_2-2}{q} +
	\right. }
\\ \nonumber
	\lefteqn{ \left.
		(\beta_2+2) (\gamma-1)
		- \frac{\gamma-2}{1+q} (1+2\:\nu^w)
	\right]  \: , }
\end{eqnarray}
which with (\ref{eq:a1}), (\ref{eq-2.3.5}) and 
$\Delta M_2 = \alpha \mej - (\gamma - 1) \mig$ gives 
the generalized $\zr^F$
\begin{eqnarray}
\label{eq:a3}
	\zr^F =
%%	\lefteqn{ \frac{1}{\alpha \mej - (\gamma - 1) \mig}
	\lefteqn{ \frac{1}{\Delta M_2}
		\left( \mej \left[ \frac{2-\beta_2}{q} - 
		\alpha (\beta_2+2) -
	\right. \right. }
\\ \nonumber
	\lefteqn{ \left. \left.
		\frac{1-\alpha}{1+q} \left(1+\frac{2}{q}+2\:\nu^{\rm FAML}\right)
		\right] +
		\mig \left[ 
		\frac{\beta_2-2}{q} + 
	\right. \right. }
\\ \nonumber
	\lefteqn{ \left. \left.
		(\beta_2+2)(\gamma-1) 
		- \frac{\gamma-2}{1+q}(1+2\:\nu^w)
	\right] \right)  \: . }
\end{eqnarray}

Six parameters describe the system:
$\mig$ and $\mej$ (ideally taken from nova models), $\nu_{\rm FAML}$
and $\alpha$ (requiring the specification of a particular FAML model), 
and $\gamma$ (or $\eta^w$), $\nu^w$ (characterizing
the wind loss during accretion). 
%If the latter two are not constant 
%due to the variable mass transfer rate, a suitable mean is required 
%to get an averaged continuous description represented by $\zr^F$.

The ratio of mass retained by the WD to that lost by the secondary
\begin{equation} 
\label{eq:a4}
 \eta = - \left( \frac{\Delta M_1}{\Delta M_2} \right)_{\rm total} = 
	\frac{ \mej - \mig }
	     { \alpha \mej - (\gamma-1) \mig }
	\: ,
\end{equation}
which relates $\mej$ and $\mig$ similar to the simple case $\eta^n$ in the
main body of the paper, now depends 
on $\alpha$ and $\gamma$.
Using a weighted $\nu$ to describe the average specific angular momentum 
gives
\begin{equation} 
\label{eq:a5}
 \nu = \frac{(\alpha-1) \mej \left( \frac{1}{q}+\nu_{\rm FAML} \right) 
	- (\gamma-2) \mig \nu^w}{ (\alpha-1) \mej - (\gamma-2) \mig } 
	\: ,
\end{equation}
and inserting these values into Eq.~(\ref{eq-2.1.14}) would also have 
directly led to Eq.~(\ref{eq:a3}).

\end{appendix}

%=============================================================================
\label{lastpage}

\begin{thebibliography}{}

\bibitem[1992]{dekool}
de Kool M.~1992, A\&A, 261, 188

\bibitem[1987]{duerbeck}
Duerbeck H.W.~1987, Space Sci.\ Rev.\. 45, 1 \& 2

\bibitem[1982]{fujimoto}
Fujimoto M.Y.~1982, ApJ, 257, 752

\bibitem[1989]{hjellming}
Hjellming M.S.~1989, PhD Thesis, Urbana--Champaign, Illinois

\bibitem[1997]{howell:etal}
Howell S., Rappaport S., Politano M.~1997, MNRAS, 287, 929

\bibitem[1983]{kato}
Kato M.~1983, PASJ, 35, 507

\bibitem[1994]{kato:hachisu}
Kato M., Hachisu I.~1994, ApJ, 437, 802

\bibitem[1988]{king}
King A.R.~1988, QJRAS, 29, 1

\bibitem[1995]{king:kolb}
King A.R., Kolb U.~1995, ApJ, 439, 330

\bibitem[1995]{king:etal}
King A.R., Kolb U., Frank J., Ritter H.~1995, ApJ, 444, L37

\bibitem[1996]{king:etal2}
King A.R., Frank J., Kolb U., Ritter H.~1996, ApJ, 467, 761

\bibitem[1995]{kley:etal}
Kley W., Shankar A., Burkert A.~1995, A\&A, 297, 739

\bibitem[1995]{kolb2}
Kolb U.~1995, in 
Bianchini A., Della~Valle M., Orio M., eds, 
Astrophysics and Space Science Library, Vol.~205, 
{\em Cataclysmic Variables}.
Kluwer Academic Publishers, Dordrecht, 
p.~511

\bibitem[1996]{kolb3}
Kolb U.~1996, in 
Evans A., Wood J.H., eds, 
IAU Coll.~158,
{\em Cataclysmic Variables and Related Objects}.
Kluwer Academic Publishers, Dordrecht, 
p.~433 

\bibitem[1992]{kolb:ritter2}
Kolb U., Ritter H.~1992, A\&A, 254, 213

\bibitem[1994]{livio}
Livio M.~1994, in 
Nussbaumer H., Orr A., eds, 
Proc.\ of the 22nd Saas Fee Advanced Course, 
{\em Interacting Binaries}.
Springer, Berlin, 
p.~135

\bibitem[1991]{livio:etal}
Livio M., Govarie A., Ritter H.~1991, A\&A, 246, 84

\bibitem[1997]{lloyd:etal}
Lloyd H.M., O'Brien T.J., Bode M.F.~1997, MNRAS, 284, 137

\bibitem[1980]{macdonald}
MacDonald J.~1980, MNRAS, 191, 933

\bibitem[1986]{macdonald2}
MacDonald J.~1986, ApJ, 305, 251

\bibitem[1989]{mazzitelli}
Mazzitelli I.~1989, ApJ, 340, 249

\bibitem[1971]{paczynski}
Paczy\'{n}ski B.~1971, ARA\&A, 9, 183

\bibitem[1984]{patterson}
Patterson J.~1984, ApJS, 54, 443

\bibitem[1996]{politano}
Politano M.~1996, ApJ, 465, 338

\bibitem[1986]{prialnik}
Prialnik D.~1986, ApJ, 310, 222

\bibitem[1995]{prialnik:kovetz}
Prialnik D., Kovetz A.~1995, ApJ, 445, 789

\bibitem[1983]{rappaport:etal}
Rappaport S., Verbunt F., Joss P.C.~1983, ApJ, 275, 713

\bibitem[1988]{ritter}
Ritter H.~1988, A\&A, 202, 93

\bibitem[1990]{ritter2}
Ritter H.~1990, in 
Cassatella A., Viotti R., eds, 
IAU Colloquium 122, 
{\em Physics of Classical Novae}.
Springer, Berlin, 
p.~313 

\bibitem[1996]{ritter3}
Ritter H.~1996, in 
Wijers R.A.M.J., Davies M.B., Tout C.A., eds, 
NATO ASI, Series C, Vol.~477, 
{\em Evolutionary Processes in Binary Stars}.
Kluwer, Dordrecht, 
p.~223 

\bibitem[1995]{ritter:kolb}
Ritter H., Kolb U. 1995, in 
Lewin W.H.G., van~Paradijs J., van~den~Heuvel E.P.J., eds, 
{\em X-ray Binaries}.
Cambridge University Press, Cambridge, 
p.~578

\bibitem[1995]{ruffert}
Ruffert M.~1995, A\&A, 311, 817

\bibitem[1989]{shara}
Shara M.M.~1989, PASP, 101, 5

\bibitem[1986]{shara:etal}
Shara M.M., Livio M., Moffat A.F.J., Orio M.~1986, ApJ, 311, 163

\bibitem[1985]{shima:etal}
Shima E., Matsuda T., Takeda H., Sawada K.~1985, MNRAS, 217, 367

\bibitem[1996]{sproats:etal}
Sproats L.N., Howell S.B., Mason K.O.~1996, MNRAS, 282, 1211

\bibitem[1983]{spruit:ritter}
Spruit H.C., Ritter H.~1983, A\&A, 124, 267

\bibitem[1993]{stehle}
Stehle R.~1993, {\em Diploma thesis},
Lud\-wig--Maxi\-mili\-ans--Universit\"at M\"unchen 

\bibitem[1996]{stehle:etal}
Stehle R., Ritter H., Kolb U.~1996, MNRAS, 279, 581 

\bibitem[1981]{verbunt:zwaan}
Verbunt F., Zwaan C.~1981, A\&A, 100, L7

\end{thebibliography}
\end{document}